\documentclass[proceedings]{JHEP3}
\usepackage{amsfonts}
\usepackage{amsmath}
\usepackage{epsfig,multicol}

\newbox\mybox

\newcommand\fverb{\setbox\mybox=\hbox\bgroup\verb}
\newcommand\fverbdo{\egroup\medskip\noindent\fbox{\unhbox\mybox}\ }
\newcommand\fverbit{\egroup\item[\fbox{\unhbox\mybox}]}
\conference{Universal boundary reflection amplitudes}
\abstract{For all affine Toda field theories we propose a new type of generic boundary 
bootstrap equations, which can be viewed as a very specific combination of 
elementary boundary bootstrap equations. These equations allow to construct
generic solutions for the boundary reflection amplitudes, which are valid for 
theories related to all simple Lie algebras, that is simply laced and non-simply laced.
We provide a detailed study of these solutions for concrete Lie algebras in various representations.}

\title{Universal boundary reflection amplitudes}
\author{Olalla Castro-Alvaredo and Andreas Fring \\
Institut f\"ur Theoretische Physik, Freie Universit\"at Berlin, \\
Arnimallee 14, D-14195 Berlin, Germany \\
E-mail: \email{olalla/fring@physik.fu-berlin.de}}

\input{tcilatex}

\begin{document}

\section{Introduction}

Similarly as in most other areas of physics, the majority of investigations
on integrable quantum field theories consists of the study of specific
examples, that is particular models. Certain general ideas and concepts can
be studied very well in this manner. However, ultimately one would like to
have formulations which go beyond particular examples as they will unravel
better which features are model dependent and which ones are of a generic
nature.

In the case of affine Toda field theory (ATFT) \cite{Toda, Toda2} such type
of formulation exists for the scattering matrices in (1+1) space-time
dimensions \cite{FKS,Oota:1997un}, where the space is a line extended
infinitely in both directions. The formulae found are of generic validity
independent of the particular algebra underlying the theory. The
understanding is not this well developed when the theory is considered in
half-space (or finite), i.e. when the line is restricted by a boundary in
one direction (or possibly both). For such theories the Yang-Baxter
equations \cite{Yang,Baxter} with reflecting boundaries have been
investigated first in \cite{Cherednik:1984vs,Sklyanin:1988yz}. Recently some
universal algebraic solutions for the Yang-Baxter equations for lattice
models have been constructed \cite{Kulish}. For a full fletched quantum
field theory one needs further properties of these solutions, such as
unitarity, crossing invariance and the bootstrap equations, which were
formulated in \cite{FK1}. The solutions for the latter system of equations
for some affine Toda field theories were first found in \cite{FK1,FK2}.
Later on, several other types of solutions for these theories have been
proposed and they have been investigated with respect to various aspects 
\cite%
{Sasaki:1993xr,FK3,Corrigan:1994ft,Kim:1995aq,Bowcock:1995vp,Corrigan:1994ft, Gandenberger:1995gg,Fujii:1995vc,Bowcock:1996gw,Penati:1996xp,Penati:1996js,Kim:1996jq,Delius:1998jw,Delius:1998rf,Dorey:1998kt, Gand1,Gand2,Bowcock1,Perkins,Riva,Ahn:2001fr,Fateev:2001rk,Delius:2001qh,Fateev1,Fateev2, Fateev3,Kojima:2002tc,Bowcock2}%
. In particular the sinh-Gordon model has attracted a considerable amount of
attention \cite{sinh,sinh5,sinh6,sinh4,sinh3,sinh1,sinh2}. Despite all this
activities, up to now closed formulae similar to the ones mentioned for the
bulk theories have not been provided for the corresponding scattering
amplitudes when boundaries are included. Furthermore, for some algebras no
solutions at all have been found yet, even on a case-by-case level. One of
the purposes of this paper is to fill in the missing gaps, but the central
aim is to supply universal, in the sense of being valid for all simple Lie
algebras and all particle types, formulae for the boundary scattering
amplitudes in affine Toda field theories.

Our manuscript is organized as follows: In section 2 we recapitulate the key
ideas of the scattering theory with reflecting boundaries and emphasize the
possibility of using certain ambiguity transformations to construct new
solutions for the boundary reflection amplitudes. Section 3 contains our
main result. We discuss here the solutions of the combined bootstrap
equations. We first recall the analogue procedure for the bulk theory and
thereafter adapt it to the situation with reflecting boundaries. We provide
generic solutions for ATFT's in form of integral representations as well as
the equivalent products of hyperbolic functions. In section 4 we provide the
explicit evaluation of our generic expression for the reflection amplitudes
for some ATFT's related to some concrete Lie algebras. In section 5 we
demonstrate in detail how our solution can be used as a \textquotedblleft
seed\textquotedblright\ for the construction of other types of solutions, in
particular we show how one may obtain from our solution, which respects the
strong-weak duality in the coupling constant, a distinct solution in which
this symmetry is broken. We provide a brief argument on how within the
bootstrap context free parameters enter into theories related to non-simply
laced algebras as well as the sinh-Gordon model. We state our conclusions in
section 6. In an appendix we provide the details for the evaluation of the
inverse q-deformed Cartan matrix and the kernel entering the integral
representation of the reflection amplitudes.

\section{Scattering theory with reflecting boundaries}

We briefly recall some well known results in order to fix our notation and
to state the problem. Exploiting the fact that the scattering of integrable
theories in 1+1 dimensions is factorized, one may formulate the theory with
the help of particle creation (annihilation) operators for the particle of
type $i$ moving with rapidity $\theta $, say $Z_{i}(\theta )$, and a
boundary in the state $\alpha $, referred to as $Z_{\alpha }$. Throughout
this paper we denote particle types and boundary degrees of freedom by Latin
and Greek letters, respectively. The operators are assumed to obey certain
exchange relations, the so-called (extended) Zamolodchikov algebra, 
\begin{eqnarray}
Z_{i}\left( \theta _{1}\right) Z_{j}\left( \theta _{2}\right)
&=&S_{ij}(\theta _{12})Z_{j}\left( \theta _{2}\right) Z_{i}\left( \theta
_{1}\right) ,  \label{ZA1} \\
Z_{i}\left( \theta \right) Z_{\alpha } &=&R_{i\alpha }\left( \theta \right)
Z_{i}\left( -\theta \right) Z_{\alpha }~.  \label{ZA2}
\end{eqnarray}%
We restrict our attention here to diagonal theories, i.e. absence of
backscattering, and do not distinguish whether we have left or right
half-spaces, i.e. if the particle hits the boundary from the left or right.
This means we assume parity invariance. We abbreviate as usual $\theta
_{12}=\theta _{1}-\theta _{2}$. The equation (\ref{ZA2}) expresses the fact
that the particle $i$ is reflected off the boundary by picking up a boundary
reflection amplitude $R$, is changing its sign of the momentum and of course
that the particle always has to stay on one particular side of the boundary.
The amplitudes obey the crossing and unitarity equations \cite{S13, ZZ, S28,
GhoshZ} 
\begin{eqnarray}
S_{ij}(\theta )S_{ji}(-\theta ) &=&1,\quad ~\ \qquad S_{i\bar{\jmath}%
}(\theta )=S_{ji}(i\pi -\theta ),~  \label{cuS} \\
R_{i\alpha }(\theta )R_{i\alpha }(-\theta ) &=&1,~\quad \qquad R_{i\alpha
}(\theta )R_{\bar{\imath}\alpha }(\theta +i\pi )=S_{ii}(2\theta )~.
\label{cuR}
\end{eqnarray}%
Most restrictive and specific to the particular theory under investigation
are the bootstrap equations \cite%
{Schroer:1976if,Karowski:1977th,Zamolodchikov:1977uc,FK1} 
\begin{eqnarray}
S_{lk}(\theta ) &=&S_{li}(\theta +i\eta _{ik}^{j})S_{lj}(\theta -i\eta
_{jk}^{i}),  \label{bootS} \\
R_{k\alpha }(\theta ) &=&R_{i\alpha }(\theta +i\eta _{ik}^{j})R_{j\alpha
}(\theta -i\eta _{jk}^{i})S_{ij}(2\theta +i\eta _{ik}^{j}+i\eta _{jk}^{i}),
\label{bootR}
\end{eqnarray}%
where the $\eta _{ik}^{j}\in \mathbb{R}$ are fusing angles which encode the
possibility that the process $i+j\rightarrow k$ takes place, i.e. particle $%
k $ can be formed as a bound state in the scattering process between the
particles $i$ and $j$. The amplitude $R_{i\alpha }(\theta )$ might have
single order poles and residues satisfying $-i\limfunc{Res}R(\theta )>0$, at
say $\theta =\eta _{i\alpha }^{\beta }$ which are interpreted as $i+\alpha
\rightarrow \beta $, that is the particle $i$ can cause the boundary to
change from the state $\alpha $ into the state $\beta $. This process is
encoded in a second type of boundary bootstrap equations \cite{GhoshZ} 
\begin{equation}
R_{j\beta }(\theta )=R_{j\alpha }(\theta )S_{ij}(\theta +i\eta _{i\alpha
}^{\beta })S_{ij}(\theta -i\eta _{i\alpha }^{\beta })~.  \label{boot2}
\end{equation}%
As in the bulk theory the solutions to these equations are not unique and
there are various ambiguities which can be used to construct from a known
solution $R_{i\alpha }(\theta )$ of the equations (\ref{cuR}), (\ref{bootR})
and (\ref{boot2}) a new solution $R_{i\alpha }^{\prime }(\theta )$ 
\begin{eqnarray}
R_{i\alpha }(\theta ,B) &\rightarrow &R_{i\alpha }^{\prime }(\theta ,B)=R_{%
\bar{\imath}\alpha }(\theta +i\pi ,B),  \label{am1} \\
R_{i\alpha }(\theta ,B) &\rightarrow &R_{i\alpha }^{\prime }(\theta
,B)=R_{i\alpha }(\theta ,B)\prod\nolimits_{j}S_{ij}(\theta ,B),  \label{am2}
\\
R_{i\alpha }(\theta ,B) &\rightarrow &R_{i\alpha }^{\prime }(\theta
,B,B^{\prime })=R_{i\alpha }(\theta ,B)\prod\nolimits_{j=1}^{\ell
}S_{ij}(\theta ,B^{\prime }),  \label{am3} \\
R_{i\alpha }(\theta ,B) &\rightarrow &R_{i\alpha }^{\prime }(\theta
,B^{\prime })=R_{i\alpha }(\theta ,B^{\prime })\qquad \text{if\quad\ }%
S_{ij}(\theta ,B)=S_{ij}(\theta ,B^{\prime })~.  \label{am4}
\end{eqnarray}%
It is clear that (\ref{am1}) always holds \cite{FK2} due to the fact that $%
S_{ij}=S_{\bar{\imath}\bar{\jmath}}$. The validity of (\ref{am2}) was noted
in \cite{FK3} for some values of $j$ and in general the new $R_{i\alpha
}^{\prime }(\theta )$ can be related to a boundary in a different state,
such as for instance $R_{i\beta }(\theta )$ \cite{FK3}. The possibility to
construct a new solution in the form (\ref{am3}) was pointed out in \cite%
{Sasaki:1993xr}, where $\ell $ denotes here the total amount of different
particle types in the theory. We have also stated explicitly some dependence
on the effective coupling $B$ or $B^{\prime }$, which will be most important
for what follows. The relevance of this is that we may change by means of (%
\ref{am3}) from a solution which respects a certain symmetry in the coupling
constant, such as the strong-weak duality, to one in which this symmetry is
broken. The relation (\ref{am4}) expresses the fact that once the bulk
theory respects a certain symmetry we may construct a new solution for the
boundary reflection amplitude in which this symmetry might be broken by
replacing the coupling according to the bulk symmetry.

Let us briefly comment on the status of explicit solutions to the boundary
reflection amplitude consistency equations (\ref{cuR}), (\ref{bootR}) and (%
\ref{boot2}). For the particular example of affine Toda field theory related
to simply laced algebras solutions to these equations were already
constructed in \cite{FK2}. Later on various other types of solutions have
been proposed and investigated with respect to various aspects \cite%
{Sasaki:1993xr,FK3,Corrigan:1994ft,Kim:1995aq,Bowcock:1995vp,Corrigan:1994ft, Gandenberger:1995gg,Fujii:1995vc,Bowcock:1996gw,Penati:1996xp,Penati:1996js,Kim:1996jq,Delius:1998jw,Delius:1998rf,Dorey:1998kt, Gand1,Gand2,Bowcock1,Perkins,Riva,Ahn:2001fr,Fateev:2001rk,Delius:2001qh,Fateev1,Fateev2, Fateev3,Kojima:2002tc,Bowcock2}%
. As we shall demonstrate, essentially all these solutions can be related to
each other or further solutions by means of (\ref{am1})-(\ref{am4}). With
regard to the above stated problem of finding closed solutions, not much
progress has been made in the last ten years. Closed solutions which respect
the bulk duality symmetry $B\rightarrow 2-B$ for the $A$ and $D$ series were
already found in \cite{FK2}. Therefore, these type of solutions reduce in
the strong as well as in the weak coupling limit to the same limit, such
that if one would like to construct a solution which relates two different
types of boundary conditions in these extremes, as proposed in \cite%
{Corrigan:1994ft}, one has to break the duality symmetry. In \cite{Fateev1}
Fateev proposed a conjecture of such type for all simply laced algebras in
form of an integral representation which generalizes a solution for the $A$
series of \cite{Corrigan:1994ft,Gand2}, the latter being simply related to
the original one in \cite{FK2} by the ambiguity transformations (\ref{am1})-(%
\ref{am4}). However, apart from $D_{n}^{(1)}$, the conjecture of \cite%
{Fateev1} provides in general only a solution of the crossing-unitarity
relations (\ref{cuR}). A solution for the boundary bootstrap equation (\ref%
{bootR}) is only proposed in some cases for some particular amplitudes. A
conjecture of a similar nature for some ATFT's related to some non-simply
laced algebras ($B_{n}^{(1)},C_{n}^{(1)},A_{2n}^{(2)}$) was formulated in 
\cite{Fateev3}. Here we aim to fill in the missing gaps, that is provide
solutions for the amplitudes and algebras not treated so far. Moreover
rather than just stating the solution as a conjecture, we propose a
systematic and unified derivation for all Lie algebras, which was absent so
far.

\section{Solutions of the combined bootstrap equations}

\subsection{Bulk theory}

We recall now the key idea of how a universal expression for the scattering
matrix can be constructed in the bulk theory and adapt the procedure
thereafter to the situation with reflecting boundaries. As already
mentioned, the central equations for the construction of the scattering
matrices when backscattering is absent are the bootstrap equations (\ref%
{bootS}). These equations express a consequence of integrability, namely
that when two particles ($i$ and $j$) fuse to a third ($k$), it is
equivalent to scatter with an additional particle ($l$) either with the two
particles before the fusing takes place or with the resulting particle after
the fusing process has happened. In principle, all these \textquotedblleft
basic\textquotedblright\ bootstrap equations (\ref{bootS}), together with
the constraints of crossing and unitarity (\ref{cuS}), are sufficient to
construct solutions for the scattering amplitudes. Proceeding this way is in
general a quite laborious task when carried out for each algebra
individually. However, in \cite{FKS} it was noted that for affine Toda field
theories there is one very special set of equations which may be obtained by
substituting the previously mentioned \textquotedblleft
basic\textquotedblright\ bootstrap equations (\ref{bootS}) into each other
in a very particular way and which were therefore referred to as
\textquotedblleft combined bootstrap equations\textquotedblright\ 
\begin{equation}
S_{ij}\left( \theta +\eta _{i}\right) S_{ij}\left( \theta -\eta _{i}\right)
=\prod\limits_{k=1}^{\ell }\prod\limits_{n=1}^{I_{ik}}S_{jk}\left( \theta
+\theta _{ik}^{n}\right) \,.  \label{Follow}
\end{equation}%
In order to keep the writing compact, the following abbreviations will be
useful 
\begin{eqnarray}
\eta _{i}:= &&\theta _{h}+t_{i}\theta _{H},~~\qquad \qquad \qquad \qquad
\theta _{ij}^{n}:=(2n-1-I_{ij})\theta _{H},~~  \label{the} \\
\theta _{h}:= &&\frac{i\pi (2-B)}{2h}=i\pi \vartheta _{h},~\qquad \qquad
~\theta _{H}:=\frac{i\pi B}{2H}\,=i\pi \vartheta _{H}.
\end{eqnarray}%
The affine Toda field theory coupling constant $\beta $ is encoded here into
the effective coupling 
\begin{equation}
B=\frac{2H\beta ^{2}}{H\beta ^{2}+4\pi h}~.  \label{effe}
\end{equation}%
We recall that ATFT's have to be considered in terms of some dual pairs of
Lie algebras, where the classical Lagrangian related to one or the other
algebra is obtained either in the weak or strong coupling limit, where $h$
and $H$ denote the respective (generalized) Coxeter numbers. The integers $%
t_{i}$ symmetrize the incidence matrix $I$, i.e. $I_{ij}t_{j}=I_{ji}t_{i}\,$%
\ and are either $t_{i}=1$ or equal to the ratio of the length of long and
short roots $t_{i}=\alpha _{i}^{2}/\alpha _{s}^{2}$, with $\alpha _{s}$
being a short root. For more details on the notation and the physics of
these models see \cite{FKS} and references therein.

The remarkable fact about equation (\ref{Follow}) is that it contains the
information about the entire bulk scattering theory. Just by solving these
equations \cite{FKS} one may derive universal expressions for the scattering
amplitudes for \emph{all} particle types $i$,$j$ and \emph{all} simple Lie
algebras. In form of an integral representation the solutions acquire a
particularly compact and neat form 
\begin{equation}
S_{ij}(\theta ,B)=\,\exp \int\limits_{0}^{\infty }\frac{dt}{t}\,\,\Phi
_{ij}(t)\sinh \left( \frac{\theta t}{i\pi }\right) \,\,,  \label{white}
\end{equation}
with 
\begin{eqnarray}
\Phi _{ij}\left( t\right) &=&8\sinh (\vartheta _{h}t)\sinh (t_{i}\vartheta
_{H}t)K_{ij}^{-1}(t)\,,  \label{rabbit} \\
K_{ij}(t)\, &=&2\cosh (\vartheta _{h}t+t_{i}\vartheta _{H}t)\delta
_{ij}-[I_{ij}]_{\bar{q}(t)}\,=(q\bar{q}^{t_{i}}+q^{-1}\bar{q}%
^{-t_{i}})\delta _{ij}-[I_{ij}]_{\bar{q}},  \label{Kq}
\end{eqnarray}
where we used the standard notation $[n]_{q}=(q^{n}-q^{-n})/(q^{1}-q^{-1})$
for q-deformed integers. The deformation parameters are related to the
coupling constant and are $q(t)=\exp (t\vartheta _{h})$ and $\bar{q}(t)=\exp
(t\vartheta _{H})$. In fact the only relevant cases here for the deformed
incidence matrix are $[0]_{\bar{q}(t)}=0$, $[1]_{\bar{q}(t)}=1$, $[2]_{\bar{q%
}(t)}=2\cosh (\vartheta _{H}t)$ and $[3]_{\bar{q}(t)}=1+2\cosh (2\vartheta
_{H}t)$.

In \cite{FKS} the combined bootstrap equations (\ref{Follow}) were derived
by translating an identity in the root space of the underlying simple Lie
algebras into an expression for the scattering matrices. We present here a
much simpler heuristic argument on how to obtain (\ref{Follow}) which is
suitable for a generalization to the situation with reflecting boundaries.
For this purpose we can formally assume the following operator product
identity

\begin{equation}
Z_{i}\left( \theta +\eta _{i}\right) Z_{i}\left( \theta -\eta _{i}\right)
=\prod\limits_{k=1}^{\ell }\prod\limits_{n=1}^{I_{ik}}Z_{k}\left( \theta
+\theta _{ik}^{n}\right) ~.  \label{Zam}
\end{equation}
It is then clear that the combined bootstrap equations (\ref{Follow}) follow
immediately when we act on both sides of (\ref{Zam}) with $Z_{j}\left(
\theta ^{\prime }\right) $ from the right (left) and move it to the left
(right) subject to the exchange relations (\ref{ZA1}). As such, this is a
rather evident statement, but the relation (\ref{Zam}) will lead to less
obvious results when reflecting boundaries are included. Here we employ (\ref%
{Zam}) only as a very useful computational tool, but it would be very
interesting to have a deeper physical understanding of this identity as well
as of the combined bootstrap equation (\ref{Follow}). Note that for each
concrete algebra we can disentangle precisely in which way (\ref{Follow})
can be manufactured from the \textquotedblleft basic" bootstrap equations (%
\ref{bootS}), but at present we are not able to provide a general
construction scheme which achieves this in a case independent manner.

\subsection{Theory with reflecting boundaries}

Let us adapt the above arguments to the situation with reflecting
boundaries. In that case we have besides the exchange relations (\ref{ZA1})
also the relations (\ref{ZA2}) at our disposal. We act now with each product
of particle states on the left and right hand side in the identity (\ref{Zam}%
) on the boundary state $Z$ in such a way that each individual particle hits
this boundary state. For simplicity we suppress here for the time being the
explicit mentioning of the boundary degree of freedom ($Z_{\alpha
}\rightarrow Z$) and assume that the boundaries remain in the same state
during this process of subsequent bombardment with particles. Ensuring that
all particles have contact with the boundary and considering thereafter the
resulting state, amounts to saying that an asymptotic in-state is related to
an out-state by a complete reversal of all signs in the momenta. Viewing
then the asymptotic states obtained in this manner as equivalent, we derive
a set of \textquotedblleft combined boundary bootstrap
equations\textquotedblright\ 
\begin{eqnarray}
R_{i}\left( \theta +\eta _{i}\right) R_{i}\left( \theta -\eta _{i}\right)
S_{ii}(2\theta ) &=&\prod\limits_{j=1}^{\ell
}\prod\limits_{n=1}^{I_{ij}}R_{j}\left( \theta +\theta _{ij}^{n}\right)
\prod\limits_{1\leq n<m\leq I_{ij}}S_{jj}(2\theta +\theta _{ij}^{n}+\theta
_{ij}^{m})~~  \notag \\
&&~\times \prod\limits_{1\leq j<k\leq \ell
}\prod\limits_{n=1}^{I_{ij}}\prod\limits_{m=1}^{I_{ik}}S_{jk}(2\theta
+\theta _{ij}^{n}+\theta _{ik}^{m})~.  \label{uni}
\end{eqnarray}
The occurrence of the bulk scattering matrices in (\ref{uni}) is due to the
fact that after a particle has hit the boundary a subsequent particle can
only reach the boundary when it first scatters with the particle already
returning back from the boundary, such that $S$ always depends on the sum of
the rapidities of the originally incoming particles. The product $%
\prod\nolimits_{1\leq n<m\leq I_{ij}}$ involving particles of the same type
only emerges for non-simply laced algebras. The equations (\ref{uni}) are
central for our investigations and we can regard them as the analogues of (%
\ref{Follow}). Therefore, we may expect that they contain all informations
of the boundary reflection. Let us solve them similarly as in \cite%
{FK1,FK2,FKS}, that is we take the logarithm of (\ref{uni}) and subsequently
use Fourier transforms. For this we define first 
\begin{equation}
\ln R_{j}(\theta )=\frac{1}{2\pi }\int dt~e^{it\theta }r_{j}(t)\quad \quad 
\text{and\quad \quad\ }\ln S_{kj}(2\theta )=\frac{1}{2\pi }\int
dt~e^{it\theta }s_{kj}(t)
\end{equation}
such that from (\ref{uni}) follows 
\begin{equation}
\sum\limits_{j=1}^{\ell }\left[ K_{ij}(t)r_{j}(t)-\!\!\!\!\sum\limits_{1\leq
n<m\leq I_{ij}}s_{jj}(t)e^{\frac{\theta _{ij}^{n}+\theta _{ij}^{m}}{2}}%
\right] =\sum\limits_{1\leq j<k\leq \ell
}\sum_{n=1}^{I_{ij}}\sum_{m=1}^{I_{ik}}~s_{jk}(t)e^{\frac{\theta
_{ij}^{n}+\theta _{ik}^{m}}{2}}-s_{ii}(t).
\end{equation}
The important difference in comparison with the bulk theory is that this
equation is non-homogeneous, in the sense that besides the quantity we want
to determine, $r_{j}(t)$, it contains terms involving quantities we already
know, namely $s_{ij}(t)$. We can use this to our advantage and solve this
equation for $r_{i}(t)$, using the integral representation for the
scattering matrix (\ref{white}). Thus we obtain the main result of this
paper, namely a closed expression for the boundary reflection matrix valid
for affine Toda field theories related to all simple Lie algebras 
\begin{equation}
\tilde{R}_{j}(\theta ,B)=\,\exp \int\limits_{0}^{\infty }\frac{dt}{t}%
\,\,\rho _{j}(t)\sinh \left( \frac{\theta t}{i\pi }\right) ,  \label{Matrix}
\end{equation}
with kernel 
\begin{eqnarray}
\rho _{i}(t) &=&\frac{1}{2}\sum\limits_{j,k,p=1}^{\ell }\left[ K^{-1}(t)%
\right] _{ij}\,\chi _{j}^{kp}(t)\Phi _{kp}\left( t/2\right) ,  \label{rho} \\
\chi _{j}^{kp} &=&(1-\delta _{pk})[I_{jk}]_{\bar{q}^{1/2}}[I_{jp}]_{\bar{q}%
^{1/2}}~-2\delta _{jk}\delta _{jp}+2\sum\limits_{n=1}^{I_{jk}-1}[n]_{\bar{q}%
}\delta _{kp}.  \label{xhi}
\end{eqnarray}
In the simply laced case the tensor $\chi $ reduces to 
\begin{equation}
\chi _{j}^{kp}=I_{jk}I_{jp}-\delta _{pk}I_{jp}\,~-2\delta _{jk}\delta _{jp}.
\label{xhis}
\end{equation}
In the derivation we made use of parity invariance, that is we used $%
s_{ij}(t)=s_{ji}(t)$. To the particular solution we constructed from (\ref%
{uni}) we refer from now on always as $\tilde{R}_{i}(\theta ,B)$ in order to
distinguish it from other solutions which might be obtained by means of the
ambiguities (\ref{am1})-(\ref{am4}).

\subsection{Integral representation versus blocks of hyperbolic functions}

The integral representations (\ref{Follow}) and (\ref{Matrix}) are very
useful starting points for various applications such as the computations of
form factors or the thermodynamic Bethe ansatz. However, one has to be
cautious when one analytically continues them into the complex rapidity
plane as one usually leaves the domain of convergence when one simply
carries out shifts in $\theta $. In addition, the singularity structure of
the integral representation is not directly obvious. Therefore one would
like to carry out the integrations which for the above type of integral
always yield some finite products of hyperbolic functions. A further reason
why we wish to carry out the integrals is that already many case-by-case
solutions for the above theories exist in the literature, which we want to
compare with.

When performing the integration, the scattering matrix of affine Toda field
theory (\ref{white}) may be represented in the form \cite{FKS} 
\begin{equation}
S_{ij}(\theta )=\prod\limits_{x=1}^{h}\prod\limits_{y=1}^{H}\left\{
x,y\right\} _{\theta }^{2\mu _{ij}(x,y)}~,  \label{SB}
\end{equation}
where 
\begin{equation}
\left\{ x,y\right\} _{\theta }:=\frac{\left[ x,y\right] _{\theta }}{\left[
x,y\right] _{-\theta }}=\exp \int\limits_{0}^{\infty }\frac{dt}{t\sinh t}%
\,\,f_{x,y}^{h,H}(t)\sinh \left( \frac{\theta t}{i\pi }\right) ,
\label{blocks}
\end{equation}
with

\begin{eqnarray}
\lbrack x,y]_{\theta }:= &&\frac{\sinh \frac{1}{2}\left[ \theta +(x-1)\theta
_{h}+(y-1)\theta _{H}\right] \sinh \frac{1}{2}\left[ \theta +(x+1)\theta
_{h}+(y+1)\theta _{H}\right] }{\sinh \frac{1}{2}\left[ \theta +(x-1)\theta
_{h}+(y+1)\theta _{H}\right] \sinh \frac{1}{2}\left[ \theta +(x+1)\theta
_{h}+(y-1)\theta _{H}\right] },~~~~~~~~~ \\
\quad \,f_{x,y}^{h,H}(t) &=&8\sinh \left( \vartheta _{h}t\right) \sinh
\left( \vartheta _{H}t\right) \sinh \left( t-x\vartheta _{h}t-y\vartheta
_{H}t\right) ~.
\end{eqnarray}%
The powers $\mu _{ij}(x,y)$ are semi-integers, which can be computed in
general from some inner products between roots and weights rotated by some
q-deformed Coxeter element \cite{Oota:1997un,FKS}\footnote{%
As is known for more than ten years, in the special case of simply laced Lie
algebras one can use the simpler formulation in terms of ordinary Coxeter
elements \cite{Dorey:1991xa,Fring:1992gh}. However, none of the formulations
will be used here.}. Alternatively, one can determine them also from the
generating function 
\begin{equation}
M_{ij}(q,\bar{q})=\sum_{x=1}^{2h}\sum_{y=1}^{2H}\mu _{ij}(x,y)q^{x}\bar{q}%
^{y}=\frac{1-q^{2h}\bar{q}^{2H}}{2}K_{ij}^{-1}(t)\,[t_{j}]_{\bar{q}}\,\,\,.
\label{M}
\end{equation}%
For this we have to view $K_{ij}^{-1}(t)$ in the q-deformed formulation (\ref%
{Kq}) and expand the right hand side of (\ref{M}) into a polynomial in $q$
and $\bar{q}$. For simply laced theories one could use simpler functions as
in that case the two dual algebras coincide, such that $h=H$ and $\left\{
x,x\right\} _{\theta }=:\left\{ x\right\} _{\theta }$. The advantage of the
formulation (\ref{M}) is that it allows for a unified treatment of all
algebras.

We can proceed now similarly for the reflection amplitudes and seek to
represent them in the form 
\begin{equation}
\tilde{R}_{i}(\theta
)=\prod\limits_{x=1}^{2h}\prod\limits_{y=1}^{2H}\left\Vert x,y\right\Vert
_{\theta }^{2\bar{\mu}_{i}(x,y)}~,  \label{RB}
\end{equation}%
where 
\begin{equation}
\left\Vert x,y\right\Vert _{\theta }:=\frac{\left\langle x,y\right\rangle
_{\theta }}{\left\langle x,y\right\rangle _{-\theta }}=\exp
\int\limits_{0}^{\infty }\frac{dt}{t\sinh t}\,\,\bar{f}_{x,y}^{h,H}(t)\sinh
\left( \frac{\theta t}{i\pi }\right) ,  \label{blockw}
\end{equation}%
with 
\begin{eqnarray}
\left\langle x,y\right\rangle _{\theta }:= &&\frac{\sinh \frac{1}{2}\left[
\theta +\frac{x-1}{2}\theta _{h}+\frac{y-1}{2}\theta _{H}\right] \sinh \frac{%
1}{2}\left[ \theta +\frac{x+1}{2}\theta _{h}+\frac{y+1}{2}\theta _{H}\right] 
}{\sinh \frac{1}{2}\left[ \theta +\frac{x-1}{2}\theta _{h}+\frac{y+1}{2}%
\theta _{H}\right] \sinh \frac{1}{2}\left[ \theta +\frac{x+1}{2}\theta _{h}+%
\frac{y-1}{2}\theta _{H}\right] },~~~~~~~ \\
\quad \,\bar{f}_{x,y}^{h,H}(t) &=&8\sinh \left( \vartheta _{h}t/2\right)
\sinh \left( \vartheta _{H}t/2\right) \sinh \left( t-x\vartheta
_{h}t/2-y\vartheta _{H}t/2\right) ~.
\end{eqnarray}%
In this case we deduce the semi-integers $\bar{\mu}_{i}(x,y)$ from 
\begin{equation}
\bar{M}_{i}(q,\bar{q})=\sum_{x=1}^{2h}\sum_{y=1}^{2H}\bar{\mu}_{i}(x,y)q^{%
\frac{x}{2}}\bar{q}^{\frac{y}{2}}=\frac{1-q^{2h}\bar{q}^{2H}}{2}\left[
K^{-1}(t)\right] _{ij}\,\chi _{j}^{kp}\left[ K^{-1}(t/2)\right]
_{kp}[t_{p}]_{\bar{q}^{1/2}}\,\,\,.  \label{Mbar}
\end{equation}%
Once again for the simply laced cases this becomes easier $\left\Vert
x,x\right\Vert _{\theta }=:\left\Vert x\right\Vert _{\theta }$, which equal
the blocks $\mathcal{W}_{h-x}(\theta )$ used in \cite{FK2}. For the
non-simply laced cases we have in principle two possible algebras, whose Lie
algebraic properties we can relate to. We make here the choice to express
everything in terms of the non-twisted algebra. Clearly one can also
formulate equivalently a generating function in terms of its dual as carried
out for the bulk theory in \cite{FKS}, but as this does not yield new
physical information, we shall be content here to do so for one algebra only.

In the following, we also abbreviate some products of the above blocks in a
more compact form 
\begin{equation}
\left\{ x,y_{n}\right\} _{\theta }:=\prod\limits_{l=0}^{n-1}\left\{
x,y+2l\right\} _{\theta },\qquad \left\Vert x,y_{n}\right\Vert _{\theta
}:=\prod\limits_{l=0}^{n-1}\left\Vert x,y+2l\right\Vert _{\theta },
\end{equation}%
and 
\begin{eqnarray}
\left\{ x_{1},y_{1}^{\mu _{1}};x_{2},y_{2}^{\mu _{2}};\cdots
;x_{n},y_{n}^{\mu _{n}}\right\} _{\theta } &:=&\left\{ x_{1},y_{1}\right\}
_{\theta }^{\mu _{1}}\left\{ x_{2},y_{2}\right\} _{\theta }^{\mu _{2}}\ldots
\left\{ x_{n},y_{n}\right\} _{\theta }^{\mu _{n}}, \\
\left\Vert x_{1},y_{1}^{\mu _{1}};x_{2},y_{2}^{\mu _{2}};\cdots
;x_{n},y_{n}^{\mu _{n}}\right\Vert _{\theta } &:=&\left\Vert
x_{1},y_{1}\right\Vert _{\theta }^{\mu _{1}}\left\Vert
x_{2},y_{2}\right\Vert _{\theta }^{\mu _{2}}\ldots \left\Vert
x_{n},y_{n}\right\Vert _{\theta }^{\mu _{n}}~.
\end{eqnarray}

\noindent For completeness we also introduce here a more elementary block
which will be useful \ for the comparison with results in the literature 
\begin{equation}
(x)_{\theta }:=\frac{\sinh (\theta +i\pi x/h)/2}{\sinh (\theta -i\pi x/h)/2}%
=-\exp \left( 2\int\limits_{0}^{\infty }\frac{dt}{t\sinh t}\,\,\sinh
t(1-x/h)\sinh \frac{\theta t}{i\pi }\right) .~
\end{equation}%
We shall also use below the blocks%
\begin{eqnarray}
\widehat{\left\Vert x\right\Vert }_{\theta } &:&=\frac{(\frac{x-1}{2})(\frac{%
x+1}{2}-h)}{(\frac{x-1+B}{2}-h)(\frac{x+1-B}{2})},  \label{bl1} \\
\overline{\left\Vert x\right\Vert }_{\theta } &:&=\frac{(\frac{h+x-1}{2})(%
\frac{h-x+1}{2})(\frac{h+x-1+B}{2})(\frac{h-x+1-B}{2})}{(\frac{h+x+1}{2})(%
\frac{h-x-1}{2})(\frac{h+x+1-B}{2})(\frac{h-x-1+B}{2})},  \label{bl2}
\end{eqnarray}%
which break the strong weak-duality.

By evaluating (\ref{Mbar}), we can determine case-by-case the powers in (\ref%
{RB}). For the simply laced case, it will turn out that our solutions
coincide with the ones found by Kim \cite{Kim:1995cf} upon the use of the
ambiguity (\ref{am1})\footnote{%
We are grateful to J.D. Kim for informing us that hep-th/9506031 v2 is
published in \cite{Kim:1995cf} and that there is some discrepancy between
the two versions.}. For the non-simply laced cases only two specific
examples have been treated in \cite{Kim}. On further solutions related to
non-simply laced algebras we shall comment below.

\section{$\tilde{R}_{i}(\protect\theta ,B)$ case-by-case}

We shall now be more concrete and evaluate our generic solution $\tilde{R}%
_{i}(\theta ,B)$ in more detail for some specified Lie algebras. We compare
with some solutions previously found in the literature. As our solutions are
invariant under the strong-weak duality transformation we commence by
comparing with those being of this type also. Apart from the $A_{2}^{(1)}$%
-case, we postpone the comparison with other types of solutions to section 5.

For the simply laced algebras the closed solution (\ref{Matrix}) admits an
even simpler general block formulation%
\begin{equation}
\tilde{R}_{i}(\theta +i\pi ,B)=\prod\limits_{x=1}^{h}\left\Vert
2x-1\right\Vert _{\theta }^{\kappa _{i}}\left( \prod\limits_{x\in \mathcal{%
\tilde{X}}_{i}}\left\Vert x\right\Vert _{\theta }\left\Vert x-2h\right\Vert
_{\theta }\right) ~,  \label{rtsim}
\end{equation}%
where the integers $\kappa _{i}$ are defined through the relation $%
\prod\nolimits_{j=1}^{\ell }S_{ij}(\theta )=\prod\nolimits_{x=1}^{h}\left\{
x\right\} _{\theta }^{\kappa _{i}}$ and the sets $\mathcal{\tilde{X}}_{i}$
are specific to each algebra. At present we do not know how a general case
independent formula which determines the sets $\mathcal{\tilde{X}}_{i}$.

\subsection{A$_{\ell }^{(1)}$-affine Toda field theory}

\subsubsection{$A_{2}^{(1)}$-affine Toda field theory}

Let us exemplify the working of the above formulae with some easy example.
As the sinh-Gordon model ($A_{1}^{(1)}$-ATFT) is very special \cite%
{sinh,sinh5,sinh6,sinh4,sinh3,sinh1,sinh2} and exhibits a distinguished
behaviour from all other ATFT's related to simply laced Lie algebras, we
consider the next simple case, namely $A_{2}^{(1)}$-ATFT. This was already
studied in \cite{FK1,FK2,Corrigan:1994ft} and especially detailed in \cite%
{Gand1}. The Coxeter number is $h=3$ in this case. The essential Lie
algebraic input here is the inverse of the q-deformed Cartan matrix (\ref{Kq}%
) 
\begin{equation}
K^{-1}(t)=\frac{1}{1+2\cosh 2t/h}\left( 
\begin{array}{cc}
2\cosh t/h & 1 \\ 
1 & 2\cosh t/h%
\end{array}%
\right) ~.
\end{equation}%
With this we compute from (\ref{rho}) and (\ref{xhis}) 
\begin{equation}
\rho _{1}(t)=\rho _{2}(t)=16\frac{\sinh [(B-2)t/12]\sinh (Bt/12)\cosh (t/6)}{%
1+2\cosh (2t/3)},
\end{equation}%
and (\ref{Mbar}) yields 
\begin{eqnarray}
\tilde{R}_{1}(\theta ,B) &=&\tilde{R}_{2}(\theta ,B)=\tilde{R}_{1}(\theta
,2-B)=\left\Vert 7,7\right\Vert _{\theta }\left\Vert 9,9\right\Vert _{\theta
}, \\
&=&-(-1)_{\theta }(-2)_{\theta }^{2}(1+B/2)_{\theta }(3-B/2)_{\theta
}(B/2+2)_{\theta }(2-B/2)_{\theta }~.
\end{eqnarray}%
We compare now with various solutions constructed before in the literature
and demonstrate that they can all be related to our solution $\tilde{R}$ by
means of the ambiguities (\ref{am1})-(\ref{am4}). We can drop the subscripts
and use $R_{1}=R_{2}=R$. In \cite{Gand1} the following solutions were
studied in detail 
\begin{eqnarray}
R^{\text{Neu}}(\theta ,B) &=&R^{\text{Neu}}(\theta ,B-2-2h)=R^{++}(\theta
,2-B)=(-2)_{\theta }(-B/2)_{\theta }(2+B/2)_{\theta }, \\
R^{--}(\theta ,B) &=&R^{--}(\theta ,B-2-2h)=-(-1)_{\theta }(B/2-1)_{\theta
}(3-B/2)_{\theta }, \\
R^{++}(\theta ,B) &=&R^{++}(\theta ,B-2-2h)=R^{\text{Neu}}(\theta
,2-B)=(-2)_{\theta }(B/2-1)_{\theta }(3-B/2)_{\theta }.\quad \quad
\end{eqnarray}%
The solution $R^{\text{Neu}}(\theta ,B)$ was already found in \cite{FK1} and
several arguments were provided in \cite{Gand1} to identify it with the
Neumann boundary condition. In addition, $R^{++}(\theta ,B)$ was related to
the fixed boundary condition. For $R^{--}(\theta ,B)$ doubts on a conclusive
identification were raised. Using now the expressions for the scattering
matrix \cite{Arinshtein:1979pb} 
\begin{eqnarray}
S_{11}(\theta ,B) &=&S_{22}(\theta ,B)=(2)_{\theta }(B-2)_{\theta
}(-B)_{\theta }, \\
S_{12}(\theta ,B) &=&S_{21}(\theta ,B)=-(1)_{\theta }(3+B)_{\theta
}(-1-B)_{\theta },
\end{eqnarray}%
it is easy to see that our solution $\tilde{R}$ is relatable to the above
ones 
\begin{eqnarray}
R^{\text{Neu}}(\theta ,B) &=&\tilde{R}(\theta ,B)S_{11}(\theta
,B/2)S_{12}(\theta ,B/2), \\
R^{--}(\theta ,B) &=&\tilde{R}(\theta +i\pi ,B)/S_{11}(\theta
,B/2)/S_{12}(\theta ,B/2), \\
R^{++}(\theta ,B) &=&\tilde{R}(\theta ,B)S_{11}(\theta ,1-B/2)S_{12}(\theta
,1-B/2).
\end{eqnarray}%
Thus we have changed by means of some ambiguities from a solution which
respects the strong-weak duality transformation $B\rightarrow 2-B$ to one in
which this symmetry is broken and replaced by the new symmetry $B\rightarrow
B-2-2h$. The solution investigated in \cite{Kim:1995cf} is related to our
solution by (\ref{am2}) 
\begin{equation}
R^{\text{K}}(\theta ,B)=\tilde{R}(\theta +i\pi ,B)~.  \label{KimCF}
\end{equation}%
For all amplitudes which were computed in \cite{Kim:1995cf} related to
simply laced Lie algebras, the relation (\ref{KimCF}) always holds.

\subsubsection{Generic A$_{\ell }^{(1)}$-affine Toda field theory}

\noindent We label the particles according to the Dynkin diagram:\bigskip

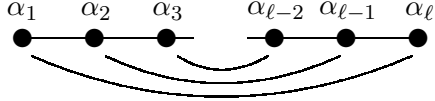
\unitlength=0.680000pt 
\begin{picture}(437.92,75.00)(50.00,95.00)
\qbezier(260.00,140.00)(365.00,95.00)(471.00,140.00)
\qbezier(301.00,140.00)(365.00,110.00)(431.00,140.00)
\qbezier(341.00,140.00)(365.00,125.00)(391.00,140.00)
\put(396.00,165.00){\makebox(0.00,0.00){${\alpha}_{\ell-2}$}}
\put(436.00,165.00){\makebox(0.00,0.00){${\alpha}_{\ell-1}$}}
\put(476.00,165.00){\makebox(0.00,0.00){${\alpha}_{\ell}$}}
\put(336.00,165.00){\makebox(0.00,0.00){${\alpha}_3$}}
\put(296.00,165.00){\makebox(0.00,0.00){${\alpha}_2$}}
\put(255.00,165.00){\makebox(0.00,0.00){${\alpha}_1$}}
\put(400.00,150.00){\line(1,0){30.00}}
\put(440.00,150.00){\line(1,0){30.00}}
\put(390.00,150.00){\line(-1,0){10.00}}
\put(340.00,150.00){\line(1,0){10.00}}
\put(300.00,150.00){\line(1,0){30.00}}
\put(260.00,150.00){\line(1,0){30.00}}
\put(395.00,150.00){\circle*{10.00}}
\put(435.00,150.00){\circle*{10.00}}
\put(475.00,150.00){\circle*{10.00}}
\put(255.00,150.00){\circle*{10.00}}
\put(295.00,150.00){\circle*{10.00}}
\put(335.00,150.00){\circle*{10.00}}
\end{picture}

\noindent The Coxeter number is $h=\ell +1$ in this case. We indicated also
the automorphism which relates the particles of type $j$ to their
anti-particles $h-j$. From the formulae derived in the appendix A.1, we
compute now the kernel of the integral representation (\ref{Matrix}) to 
\begin{equation}
\rho _{j}^{\mathbf{A}_{\ell }}(t)=\frac{4\sinh (\frac{2-B}{4h})t\sinh \frac{%
Bt}{4h}\cosh \frac{t}{2h}\sinh (\frac{1-h}{2h})t\sinh \frac{jt}{h}\sinh (%
\frac{h-j}{h})t}{\sinh t\cosh \frac{t}{2}\sinh ^{2}\frac{t}{h}}~.
\end{equation}%
Solving the integral or more practical using the generating function (\ref%
{Mbar}), we transform this into the block representation (\ref{RB}) and find 
\begin{equation}
\tilde{R}_{j}(\theta +i\pi ,B)=\tilde{R}_{h-j}(\theta +i\pi
,B)=\prod\limits_{p=1}^{j}\prod\limits_{k=p}^{h-p}\left\Vert 2k-1\right\Vert
_{\theta }\quad \text{for~}j\leq h/2~.  \label{an}
\end{equation}%
We used here the well-known relation between particles and anti-particles
indicated above. For $j=1$ our solution coincides with the amplitude found
in \cite{Kim:1995cf} shifted by $i\pi $ in the rapidity. More solutions were
not reported in \cite{Kim:1995cf} for this algebra.

Computing $\prod\nolimits_{j=1}^{\ell }S_{ij}(\theta
)=\prod\nolimits_{p=1}^{i}\prod\nolimits_{k=p}^{h-p}\left\{ k\right\}
_{\theta }$, we note here the additional structure (\ref{rtsim}) with $%
\mathcal{\tilde{X}}_{i}$ $=\emptyset $ for $1\leq i\leq \ell $.

\subsection{D$_{\ell }^{(1)}$-affine Toda field theory}

We proceed now similarly and label the particles according to the Dynkin
diagram\bigskip

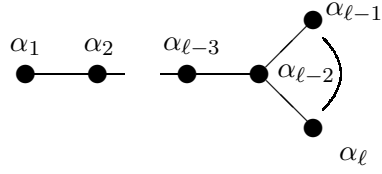
\unitlength=0.680000pt 
\begin{picture}(437.92,130.00)(50.00,65.00)
\qbezier(420.00,170.00)(440.00,150.00)(420.00,130.00)
\put(437.50,104.17){\makebox(0.00,0.00){$\alpha_{\ell}$}}
\put(437.92,184.17){\makebox(0.00,0.00){${\alpha}_{\ell -1}$}}
\put(411.17,149.59){\makebox(0.00,0.00){${\alpha}_{\ell-2}$}}
\put(347.50,165.00){\makebox(0.00,0.00){${\alpha}_{\ell-3}$}}
\put(296.00,165.00){\makebox(0.00,0.00){${\alpha}_2$}}
\put(255.00,165.00){\makebox(0.00,0.00){${\alpha}_1$}}
\put(389.00,146.33){\line(1,-1){22.67}}
\put(411.67,176.67){\line(-1,-1){23.00}}
\put(350.00,150.00){\line(1,0){30.00}}
\put(330.00,150.00){\line(1,0){10.00}}
\put(300.00,150.00){\line(1,0){10.00}}
\put(260.00,150.00){\line(1,0){30.00}}
\put(415.00,120.00){\circle*{10.00}}
\put(415.00,180.00){\circle*{10.00}}
\put(385.00,150.00){\circle*{10.00}}
\put(345.00,150.00){\circle*{10.00}}
\put(255.00,150.00){\circle*{10.00}}
\put(295.00,150.00){\circle*{10.00}}
\end{picture}

\noindent In the $\mathbf{D}_{\ell }$-case the Coxeter number is $h=2(\ell
-1)$. As indicated most particles are self-conjugate apart from the two
\textquotedblleft spinors" at the end which are conjugate to each other.
From the formulae derived in the appendix A.2,We compute now the kernel in (%
\ref{Matrix}) for ~$1\leq j\leq \ell -2$ to 
\begin{equation}
\rho _{j}^{\mathbf{D}_{\ell }}(t)=\frac{16\sinh (\frac{2-B}{4h})t\sinh \frac{%
Bt}{4h}\cosh \frac{(h-2)t}{4h}\sinh \frac{t}{4}\sinh \frac{(j-h)t}{2h}\sinh 
\frac{jt}{2h}}{\sinh t\sinh ^{2}\frac{t}{2h}}~\   \label{rho1}
\end{equation}%
and for the spinors 
\begin{equation}
\rho _{\ell }^{\mathbf{D}_{\ell }}(t)=\rho _{\ell -1}^{\mathbf{D}_{\ell
}}(t)=\frac{8\sinh (\frac{2-B}{4h})t\sinh \frac{Bt}{4h}\sinh \frac{(1-h)t}{2h%
}\sinh \frac{(h+1-2\left[ \ell /2\right] )t}{2h}\sinh \frac{t\left[ \ell /2%
\right] }{h}}{\sinh t\sinh \frac{t}{h}\sinh \frac{t}{2h}}.  \label{rho2}
\end{equation}%
Solving the integral in (\ref{Matrix}) or using the generating function (\ref%
{Mbar}), we find the following compact and closed expressions for the
reflection matrices in terms of hyperbolic functions 
\begin{eqnarray}
\tilde{R}_{j}(\theta +i\pi ) &=&\left[ \prod\limits_{k=1}^{j}\left\Vert
h-2k+1\right\Vert \right] \prod\limits_{p=1}^{j}\prod\limits_{k=p}^{h-p}%
\left\Vert 2k-1\right\Vert \quad \quad \text{for\quad \quad }j=1,\ldots
,\ell -2,\quad  \label{dn1} \\
\tilde{R}_{\ell }(\theta +i\pi ) &=&\tilde{R}_{\ell -1}(\theta +i\pi
)=\prod\limits_{p=1}^{\left[ \ell /2\right] }\prod\limits_{k=2p-1}^{h-2p+1}%
\left\Vert 2k-1\right\Vert .  \label{dn2}
\end{eqnarray}%
For $D_{4}^{(1)}$ our solution agrees with the one reported in \cite%
{Kim:1995qe} when shifted by $i\pi $ in the rapidity. This is one of the few
examples for which a perturbative calculation has been carried out, using
Neumann boundary conditions in this case. For higher ranks only a solution
for $j=1$ was also reported in \cite{Kim:1995cf}, which once again coincides
with ours subject to the relation (\ref{KimCF}).

Computing now $\prod\nolimits_{j=1}^{\ell }S_{ij}(\theta )_{\theta }$, we
note that $\tilde{R}$ admits the alternative form (\ref{rtsim}) with

\begin{eqnarray}
\mathcal{\tilde{X}}_{i} &=&\emptyset \qquad \text{for }i=1,\ell -1,\ell \\
\mathcal{\tilde{X}}_{i} &=&\bigcup\limits_{1\leq
k<[(2i+1)/4]}\{h+4k-2i-1\}\qquad \text{for }2\leq i\leq \ell -2~.
\end{eqnarray}

\subsection{E$_{6}^{(1)}$-affine Toda field theory}

The labeling of the particle types is now according to the Dynkin
diagram\bigskip

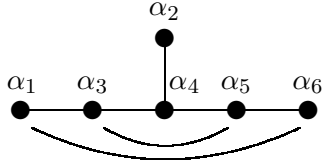
\unitlength=0.680000pt 
\begin{picture}(370.00,107.58)(0.00,-20.00)
\qbezier(211.00,28.00)(285.00,-7.00)(360.00,28.00)
\qbezier(251.00,28.00)(285.00,8.0)(320.00,28.00)
\put(365.00,52.00){\makebox(0.00,0.00){$\alpha_6$}}
\put(325.00,52.00){\makebox(0.00,0.00){$\alpha_5$}}
\put(296.25,52.17){\makebox(0.00,0.00){${\alpha}_4$}}
\put(285.00,93.00){\makebox(0.00,0.00){${\alpha}_2$}}
\put(245.00,52.00){\makebox(0.00,0.00){${\alpha}_3$}}
\put(206.00,52.00){\makebox(0.00,0.00){${\alpha}_1$}}
\put(330.00,38.00){\line(1,0){30.00}}
\put(285.33,72.67){\line(0,-1){31.00}}
\put(290.00,38.00){\line(1,0){30.00}}
\put(250.00,38.00){\line(1,0){30.00}}
\put(210.00,38.00){\line(1,0){30.00}}
\put(285.00,78.00){\circle*{10.00}}
\put(245.00,38.00){\circle*{10.00}}
\put(325.00,38.00){\circle*{10.00}}
\put(365.00,38.00){\circle*{10.00}}
\put(285.00,38.00){\circle*{10.00}}
\put(205.00,38.00){\circle*{10.00}}
\end{picture}

\noindent The Coxeter number equals $h=12$ in this case. We indicated the
conjugation properties. From the formulae derived in the appendix A.3, we
can obtain the integral representation (\ref{Matrix}) from which we deduce
the block representation (\ref{RB}) directly or use the generating function (%
\ref{Mbar}). We find 
\begin{eqnarray}
\tilde{R}_{1}(\theta +i\pi ) &=&\tilde{R}_{6}(\theta +i\pi )=\left\Vert
1;3;5;7^{2};9^{2};11^{2};13^{2};15^{2};17;19;21\right\Vert _{\theta }~,
\label{r1} \\
\tilde{R}_{3}(\theta +i\pi ) &=&\tilde{R}_{5}(\theta +i\pi )=\left\Vert
1;3^{2};5^{3};7^{4};9^{4};11^{4};13^{4};15^{3};17^{2};19^{2};21\right\Vert
_{\theta }~, \\
\tilde{R}_{2}(\theta +i\pi ) &=&\left\Vert
1;3;5^{2};7^{3};9^{3};11^{3};13^{2};15^{3};17^{2};19;21\right\Vert _{\theta
}~, \\
\tilde{R}_{4}(\theta +i\pi ) &=&\left\Vert
1;3^{3};5^{5};7^{6};9^{6};11^{6};13^{5};15^{4};17^{3};19^{2};21\right\Vert
_{\theta }~.  \label{r4}
\end{eqnarray}%
This solution coincides precisely with the amplitudes found in \cite%
{Kim:1995cf} shifted by $i\pi $ in the rapidity. We note here that the
structure of the blocks in (\ref{r1})-(\ref{r4}) can be encoded elegantly
into the form (\ref{rtsim}) with%
\begin{equation}
\mathcal{\tilde{X}}_{1}^{E_{6}}=\mathcal{\tilde{X}}_{6}^{E_{6}}=\emptyset
,\quad \mathcal{\tilde{X}}_{3}^{E_{6}}=\mathcal{\tilde{X}}%
_{5}^{E_{6}}=\{7\},\quad \mathcal{\tilde{X}}_{2}^{E_{6}}=\{11\},\quad 
\mathcal{\tilde{X}}_{4}^{E_{6}}=\{5,7,9\}.
\end{equation}

\subsection{E$_{7}^{(1)}$-affine Toda field theory}

The labeling of the particle types is now according to the Dynkin
diagram\bigskip

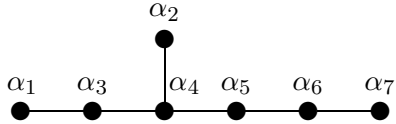
\unitlength=0.680000pt 
\begin{picture}(370.00,107.58)(0.00,0.00)
\put(405.00,52.00){\makebox(0.00,0.00){$\alpha_7$}}
\put(365.00,52.00){\makebox(0.00,0.00){$\alpha_6$}}
\put(325.00,52.00){\makebox(0.00,0.00){$\alpha_5$}}
\put(296.25,52.17){\makebox(0.00,0.00){${\alpha}_4$}}
\put(285.00,93.00){\makebox(0.00,0.00){${\alpha}_2$}}
\put(245.00,52.00){\makebox(0.00,0.00){${\alpha}_3$}}
\put(206.00,52.00){\makebox(0.00,0.00){${\alpha}_1$}}
\put(370.00,38.00){\line(1,0){30.00}}
\put(330.00,38.00){\line(1,0){30.00}}
\put(285.33,72.67){\line(0,-1){31.00}}
\put(290.00,38.00){\line(1,0){30.00}}
\put(250.00,38.00){\line(1,0){30.00}}
\put(210.00,38.00){\line(1,0){30.00}}
\put(285.00,78.00){\circle*{10.00}}
\put(245.00,38.00){\circle*{10.00}}
\put(325.00,38.00){\circle*{10.00}}
\put(365.00,38.00){\circle*{10.00}}
\put(405.00,38.00){\circle*{10.00}}
\put(285.00,38.00){\circle*{10.00}}
\put(205.00,38.00){\circle*{10.00}}
\end{picture}

\noindent The Coxeter number equals $h=18$ for $E_{7}$. All particles are
self-conjugate. Using the formulae of appendix A.4, we can again either
solve the integral (\ref{Matrix}) or use the generating function (\ref{Mbar}%
) and deduce the block representation (\ref{RB}). We find that these
amplitudes coincide precisely with those reported in \cite{Kim:1995cf}
(published version) when shifted by $i\pi $ in the rapidity. We note here
once more, that they admit the additional structure (\ref{rtsim}) with%
\begin{eqnarray}
\mathcal{\tilde{X}}_{1}^{E_{7}} &=&\{17\},\quad \mathcal{\tilde{X}}%
_{2}^{E_{7}}=\{11\},\quad \mathcal{\tilde{X}}_{3}^{E_{7}}=\{7,11,15\},\quad 
\mathcal{\tilde{X}}_{4}^{E_{7}}=\{5,7,9^{2},11,13^{2},17\},\quad \\
\mathcal{\tilde{X}}_{5}^{E_{7}} &=&\{7,9,11,15\},\quad \mathcal{\tilde{X}}%
_{6}^{E_{7}}=\{9,17\},\quad \mathcal{\tilde{X}}_{7}^{E_{7}}=\emptyset .
\end{eqnarray}

\subsection{E$_{8}^{(1)}$-affine Toda field theory}

In this case we label the particles according to the Dynkin diagram\bigskip

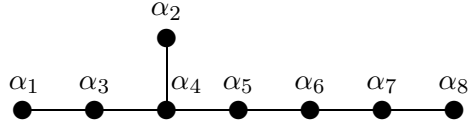
\unitlength=0.680000pt 
\begin{picture}(370.00,107.58)(0.00,0.00)
\put(445.00,52.00){\makebox(0.00,0.00){$\alpha_8$}}
\put(405.00,52.00){\makebox(0.00,0.00){$\alpha_7$}}
\put(365.00,52.00){\makebox(0.00,0.00){$\alpha_6$}}
\put(325.00,52.00){\makebox(0.00,0.00){$\alpha_5$}}
\put(296.25,52.17){\makebox(0.00,0.00){${\alpha}_4$}}
\put(285.00,93.00){\makebox(0.00,0.00){${\alpha}_2$}}
\put(245.00,52.00){\makebox(0.00,0.00){${\alpha}_3$}}
\put(206.00,52.00){\makebox(0.00,0.00){${\alpha}_1$}}
\put(410.00,38.00){\line(1,0){30.00}}
\put(370.00,38.00){\line(1,0){30.00}}
\put(330.00,38.00){\line(1,0){30.00}}
\put(285.33,72.67){\line(0,-1){31.00}}
\put(290.00,38.00){\line(1,0){30.00}}
\put(250.00,38.00){\line(1,0){30.00}}
\put(210.00,38.00){\line(1,0){30.00}}
\put(285.00,78.00){\circle*{10.00}}
\put(245.00,38.00){\circle*{10.00}}
\put(325.00,38.00){\circle*{10.00}}
\put(365.00,38.00){\circle*{10.00}}
\put(405.00,38.00){\circle*{10.00}}
\put(445.00,38.00){\circle*{10.00}}
\put(285.00,38.00){\circle*{10.00}}
\put(205.00,38.00){\circle*{10.00}}
\end{picture}

\noindent The Coxeter number equals $h=30$ for $E_{8}$. All particles are
self-conjugate. Using the formulae of appendix A.5, we can solve the
integral (\ref{Matrix}) or use the generating function (\ref{Mbar}) and
deduce the block representation (\ref{RB}). Once more we find that these
amplitudes coincide precisely with those reported in \cite{Kim:1995cf}
(published version) when shifted by $i\pi $ in the rapidity. They admit the
additional structure (\ref{rtsim}) with%
\begin{eqnarray}
\mathcal{\tilde{X}}_{1}^{E_{8}} &=&\{17,29\},\quad \mathcal{\tilde{X}}%
_{2}^{E_{8}}=\{11,15,19,23\},\quad \mathcal{\tilde{X}}_{3}^{E_{8}}=%
\{7,11,13,15,17,19^{2},23,27\},\quad \\
\mathcal{\tilde{X}}_{4}^{E_{8}}
&=&\{5,7,9^{2},11^{2},13^{3},15^{2},17^{3},19^{2},21^{3},23,25^{2},29\}, \\
\mathcal{\tilde{X}}_{5}^{E_{8}}
&=&\{7,9^{2},11^{2},13,15^{2},17,19^{2},21,23^{2},27\}, \\
\mathcal{\tilde{X}}_{6}^{E_{8}} &=&\{9,11,13,17,19,21,25,29\},\quad \mathcal{%
\tilde{X}}_{7}^{E_{8}}=\{11,19,27\},\quad \mathcal{\tilde{X}}%
_{8}^{E_{8}}=\{29\}~.
\end{eqnarray}

\subsection{(B$_{\ell }^{(1)},$A$_{2\ell -1}^{(2)}$)-affine Toda field theory%
}

As not many examples for reflection amplitudes of ATFT's related to
non-simply laced Lie algebras have been computed, we consider it useful to
start with some specific example before turning to the generic case. In
general we label the particle types according to the Dynkin diagram\medskip

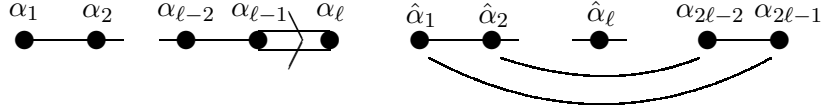
\begin{center}
{\unitlength=0.680000pt 
\begin{picture}(430.00,78.12)(0.00,0.00)
\qbezier(270.00,38.12)(330.00,18.13)(380.00,38.12)
\qbezier(230.00,38.12)(326.25,-13.12)(420.00,38.12)
\put(390.00,48.12){\line(1,0){30.00}}
\put(270.00,48.12){\line(1,0){10.00}}
\put(230.00,48.12){\line(1,0){30.00}}
\put(430.25,62.71){\makebox(0.00,0.00){$\alpha_{2\ell-1}$}}
\put(386.67,62.71){\makebox(0.00,0.00){$\alpha_{2\ell-2}$}}
\put(326.00,63.12){\makebox(0.00,0.00){$\hat{\alpha}_\ell$}}
\put(266.25,62.29){\makebox(0.00,0.00){$\hat{\alpha}_2$}}
\put(226.00,62.12){\makebox(0.00,0.00){$\hat{\alpha}_1$}}
\put(175.42,63.54){\makebox(0.00,0.00){$\alpha_\ell$}}
\put(136.00,63.12){\makebox(0.00,0.00){$\alpha_{\ell-1}$}}
\put(95.00,63.12){\makebox(0.00,0.00){$\alpha_{\ell-2}$}}
\put(46.00,63.12){\makebox(0.00,0.00){$\alpha_2$}}
\put(5.00,64.12){\makebox(0.00,0.00){$\alpha_1$}}
\put(160.00,48.12){\line(-1,-2){8.00}}
\put(160.00,48.12){\line(-1,2){8.00}}
\put(135.00,42.79){\line(1,0){40.33}}
\put(135.00,53.12){\line(1,0){39.67}}
\put(330.00,48.12){\line(1,0){10.00}}
\put(310.00,48.12){\line(1,0){10.00}}
\put(100.00,48.12){\line(1,0){30.00}}
\put(80.00,48.12){\line(1,0){10.00}}
\put(50.00,48.12){\line(1,0){10.00}}
\put(10.00,48.12){\line(1,0){30.00}}
\put(425.00,48.12){\circle*{10.00}}
\put(385.00,48.12){\circle*{10.00}}
\put(325.00,48.12){\circle*{10.00}}
\put(265.00,48.12){\circle*{10.00}}
\put(225.00,48.12){\circle*{10.00}}
\put(175.00,48.12){\circle*{10.00}}
\put(135.00,48.12){\circle*{10.00}}
\put(95.00,48.12){\circle*{10.00}}
\put(45.00,48.12){\circle*{10.00}}
\put(5.00,48.12){\circle*{10.00}}
\end{picture}}
\end{center}

\subsubsection{$(B_{2}^{(1)},A_{3}^{(2)})$-affine Toda field theory}

In this case we have for the (generalized) Coxeter numbers $h=4$, $H=6$, the
incidence matrix $I_{12}=2$, $I_{21}=1$ and the symmetrizers $t_{1}=2$ and $%
t_{2}=1$. This is already enough Lie algebraic information needed for the
computation of the relevant matrices in equation (\ref{rho}). We obtain 
\begin{equation}
K^{-1}(t)=\frac{1}{\cosh t/2}\left( 
\begin{array}{ll}
\cosh t(\vartheta _{h}+\vartheta _{H}) & \cosh t\vartheta _{H} \\ 
1/2 & \cosh t(\vartheta _{h}+2\vartheta _{H})%
\end{array}%
\right) ~.
\end{equation}%
With the help of this matrix we can evaluate the scattering amplitudes (\ref%
{white}) and (\ref{Matrix}). Alternatively we may compute the representation
in terms of blocks from this. The expression (\ref{SB}) yields 
\begin{equation}
S_{11}(\theta )=\left\{ 1,1\right\} \left\{ 1,3\right\} \left\{ 3,3\right\}
\left\{ 3,5\right\} _{\theta },~~S_{22}(\theta )=\left\{ 1,1\right\} \left\{
3,5\right\} _{\theta },~~S_{12}(\theta )=\left\{ 2,2\right\} \left\{
2,4\right\} _{\theta }
\end{equation}%
and (\ref{RB}) 
\begin{eqnarray}
\tilde{R}_{1}(\theta +i\pi ) &=&\left\Vert 1,1\right\Vert \left\Vert
1,3\right\Vert \left\Vert 3,3\right\Vert \left\Vert 3,5\right\Vert
\left\Vert 3,7\right\Vert \left\Vert 5,5\right\Vert _{\theta }~,  \label{11}
\\
\tilde{R}_{2}(\theta +i\pi ) &=&\left\Vert 1,1\right\Vert \left\Vert
3,3\right\Vert \left\Vert 3,5\right\Vert \left\Vert 5,9\right\Vert _{\theta
}~.  \label{12}
\end{eqnarray}%
The solutions (\ref{11}), (\ref{12}) correspond precisely to those found by
J.D. Kim in \cite{Kim} after re-defining the effective coupling as $%
B\rightarrow B/2$ and shifting $\theta $ by $i\pi $. These solutions are
especially trustworthy as they have also been double checked against
perturbation theory. As the non-simply laced cases are not yet covered very
much in the literature, we consider it useful to perform some more analysis
at least for this case. Let us study the bootstrap equation (\ref{boot2})
which relates different boundary states to each other in more detail.
Adopting here the same principle as in the bulk, see \cite{FKS,FK3} and
references therein, namely that $-i\limfunc{Res}R(\theta =\eta )>0$ in the
entire range of the coupling constant we find here 
\begin{equation}
-i\limfunc{Res}_{\theta \rightarrow \eta _{2\alpha }^{\beta }=\theta
_{h}+\theta _{H}}\tilde{R}_{i\alpha }(\theta +i\pi )>0~.  \label{res}
\end{equation}%
Solving for this angle $\eta _{2\alpha }^{\beta }$ the bootstrap equation (%
\ref{boot2}) yields 
\begin{equation}
R_{i\alpha }(\theta )=S_{i1}(-\theta )R_{i\beta }(\theta )~.
\end{equation}%
Considering now the new solution $R_{i\beta }(\theta )$, we observe that 
\begin{equation}
-i\limfunc{Res}_{\theta \rightarrow \eta _{2\beta }^{\alpha }=3\theta
_{h}+5\theta _{H}}\tilde{R}_{i\beta }(\theta +i\pi )>0~.
\end{equation}%
These are the only poles with the property to have positive definitive sign
in the entire range of the coupling constant, such that we have just the two
boundary states $\alpha $ and $\beta $. The corresponding energies are
computed in the same way as in \cite{GhoshZ,FK3}. Using that 
\begin{equation}
m_{1}=m\sinh (2\theta _{h}+4\theta _{H})\quad \text{and\quad }m_{2}=m\sinh
(\theta _{h}+\theta _{H})~,
\end{equation}%
with $m$ being an overall mass scale, we find for the energies of the two
boundary states 
\begin{equation}
E_{\alpha }=E_{\beta }-m_{2}\cosh (\theta _{h}+\theta _{H})=E_{\beta
}-m_{1}/2~,
\end{equation}%
such that it appears that Kim's solution is not the ground state. When
performing the same analysis for our solution $\tilde{R}_{i}(\theta )$ we
find that there is no simple order pole which respects (\ref{res}), such
that there is only one state in that case.

\subsubsection{$(B_{3}^{(1)},A_{5}^{(2)})$-affine Toda field theory}

As the previous case can be related trivially to a solution which may be
found already in the literature, let us present a case not dealt with so
far. The (generalized) Coxeter numbers for ($B_{3}^{(1)},A_{5}^{(2)}$) are $%
h=6$ and $H=10$. According to the corresponding Dynkin diagram of $%
B_{3}^{(1)}$, we have $t_{1}=t_{2}=2$ and $t_{3}=1$. We evaluate 
\begin{equation}
K^{-1}=\frac{1}{\det K}\left( 
\begin{array}{ccc}
q^{2}\bar{q}^{3}\ +q^{-2}\bar{q}^{-3} & q\bar{q}\ +q^{-1}\bar{q}^{-1} & \bar{%
q}\ +\bar{q}^{-1} \\ 
q\bar{q}\ +q^{-1}\bar{q}^{-1} & \bar{q}\ +\bar{q}^{-1}+q^{2}\bar{q}^{3}\
+q^{-2}\bar{q}^{-3} & q\bar{q}\ +q^{-1}\bar{q}^{-1}+q\bar{q}^{3}\ +q^{-1}%
\bar{q}^{-3} \\ 
1 & q\bar{q}^{2}\ +q^{-1}\bar{q}^{-2} & 1+q^{2}\bar{q}^{4}\ +q^{-2}\bar{q}%
^{-4}%
\end{array}
\right) ,~~~~~~  \label{kb33}
\end{equation}
with $\det K=q^{3}\bar{q}^{5}\ +q^{-3}\bar{q}^{-5}$. From this we obtain 
\begin{eqnarray}
\tilde{R}_{1}(\theta +i\pi ) &=&\left\|
1,1_{2};3,5_{2};5,7_{3};7,9;9,13\right\| _{\theta },  \label{b31} \\
\tilde{R}_{2}(\theta +i\pi ) &=&\left\|
1,1_{2};3,3_{4};3,5_{3};5,5_{3};7,9_{2};7,15;9,13\right\| _{\theta },
\label{b32} \\
\tilde{R}_{3}(\theta +i\pi ) &=&\left\|
1,1;3,3_{2};5,7_{2};5,9;7,11_{2};9,17\right\| _{\theta }.  \label{b33}
\end{eqnarray}
We are not aware of any solution of this type occurring in the literature
for this algebra.

\subsubsection{Generic (B$_{\ell }^{(1)},$A$_{2\ell -1}^{(2)}$)-affine Toda
field theory}

In this case we have $h=2\ell $ and $H=2(2\ell -1)$. The task of finding
general block expressions for all reflection amplitudes, such as for
instance in (\ref{an}) for $A_{\ell }^{(1)}$ turns out to be quite involved
in this case and therefore we present only closed expressions corresponding
to some specific particles. For the first particle we find for $\ell \neq 2$ 
\begin{eqnarray}
\tilde{R}_{1}^{(\mathbf{B}_{\ell }^{(1)},\mathbf{A}_{2\ell
-1}^{(2)})}(\theta +i\pi ,B) &=&\left\Vert h-1,\left[ H-3\right]
_{3};h+1,H-1\right\Vert _{\theta }\prod_{k=\ell +1}^{2(\ell -1)}\left\Vert
2k+1,\left[ 4k-3\right] _{2}\right\Vert _{\theta }  \notag \\
&&\!\!\times \prod\limits_{k=0}^{\ell -2}\left\Vert 2k+1,\left[ 4k+1\right]
_{2}\right\Vert _{\theta },
\end{eqnarray}%
whereas for the second with $\ell \neq 2,3$ we obtain 
\begin{eqnarray}
\tilde{R}_{2}^{(\mathbf{B}_{\ell }^{(1)},\mathbf{A}_{2\ell
-1}^{(2)})}(\theta +i\pi ,B) &=&\left\Vert 1,1_{2};h-3,\left[ H-7\right]
_{3};h-3,\left[ H-5\right] _{2};h-1,\left[ H-5\right] _{4};\right.  \notag \\
&&h-1,H-1;h+1,\left[ H-1\right] _{2};h+1,H+5;h+3,\left[ H+3\right] _{2}; 
\notag \\
&&\left. h+3,H+3;2h-3,\left[ 2H-7\right] _{2}\right\Vert _{\theta
}\prod\limits_{k=0}^{\ell -2}\left\Vert 2k+3,\left[ 4k+5\right]
_{2}\right\Vert _{\theta }^{2}~  \notag \\
&&\!\!\times \prod\limits_{k=\ell +1}^{2(\ell -1)}\left\Vert 2k+3,\left[ 4k+1%
\right] _{2}\right\Vert _{\theta }^{2}.
\end{eqnarray}%
For the last two particles the amplitudes are 
\begin{eqnarray}
\tilde{R}_{\ell -1}^{(\mathbf{B}_{\ell }^{(1)},\mathbf{A}_{2\ell
-1}^{(2)})}(\theta +i\pi ,B) &=&\prod\limits_{k=0}^{\ell -2}\left\Vert 2k+1,%
\left[ 4k+1\right] _{2};2k+3,\left[ 4k+3\right] _{3};4k+5,8k+5\right\Vert
_{\theta }  \notag \\
&&\times \prod\limits_{k=1}^{\ell -2}\left\Vert 2k+3,4k+5;4k+3,\left[ 8k+1%
\right] _{2};4k+3,8k+7\right\Vert _{\theta }  \notag \\
&&\times \prod_{n=0}^{\ell -4}\prod\limits_{k=\ell -n}^{2(\ell
-n-2)}\left\Vert 2k+1,\left[ 4k-3\right] _{4}\right\Vert _{\theta },
\end{eqnarray}%
\begin{equation}
\tilde{R}_{\ell }^{(\mathbf{B}_{\ell }^{(1)},\mathbf{A}_{2\ell
-1}^{(2)})}(\theta +i\pi ,B)=\left[ \prod\limits_{k=0}^{\ell -1}\left\Vert
4k+1,8k+1\right\Vert _{\theta }\right] \prod_{n=0}^{\ell -2}\prod_{k=\ell
-n}^{2(\ell -n-1)}\left\Vert 2k-1,\left[ 4k-5\right] _{2}\right\Vert
_{\theta },
\end{equation}%
In particular, one can easily specialize these functions to reproduce the
examples $\ell =2,3$ treated in the previous subsections. We also report
here for $i=\ell ,$ the corresponding integral representation which is given
in terms of the kernel, 
\begin{eqnarray}
&&\sum_{j,k,p=1}^{\ell }\left[ K^{(\mathbf{B}_{\ell }^{(1)},\mathbf{A}%
_{2\ell -1}^{(2)})}(t)\right] _{\ell j}^{-1}\chi _{j}^{kp}\left[ K^{(\mathbf{%
B}_{\ell }^{(1)},\mathbf{A}_{2\ell -1}^{(2)})}(t/2)\right] _{kp}^{-1}\left[
t_{p}\right] _{\bar{q}^{1/2}}  \notag \\
&=&\frac{4\sinh \frac{th(\vartheta _{h}+2\vartheta _{H})}{4}}{\sinh t\sinh 
\frac{t(\vartheta _{h}+2\vartheta _{H})}{2}}\left( \frac{\cosh \frac{%
t\vartheta _{H}}{2}\sinh \frac{t(\vartheta _{h}+2\vartheta _{H})(2-h)}{4}%
\sinh \frac{t(\vartheta _{h}(h-1)+2\vartheta _{H}(h-2))}{4}}{\sinh
t(\vartheta _{h}+2\vartheta _{H})}\right.  \notag \\
&&\left. +\cosh \frac{th(\vartheta _{h}+2\vartheta _{H})}{4}\sinh \frac{%
t(2(1-H)\vartheta _{H}-H\vartheta _{h})}{4}\right) .
\end{eqnarray}%
As the expressions for the other amplitudes turn out to be rather lengthy we
do not report them here, but it should be clear \ how to obtain them.

\subsection{(C$_{\ell }^{(1)},$D$_{\ell +1}^{(2)}$)-affine Toda field theory}

We label the particle types according to the Dynkin diagram\medskip

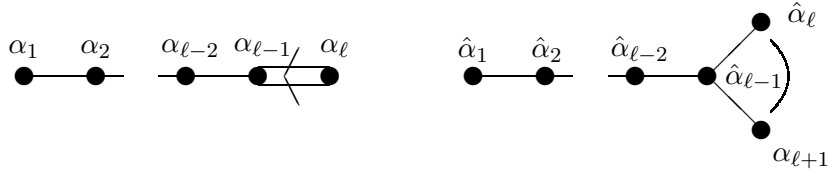
\unitlength=0.680000pt 
\begin{picture}(437.92,149.59)(-40.00,50.00)
\qbezier(420.00,170.00)(440.00,150.00)(420.00,130.00)
\put(437.50,104.17){\makebox(0.00,0.00){$\alpha_{\ell+1}$}}
\put(437.92,184.17){\makebox(0.00,0.00){$\hat{\alpha}_\ell$}}
\put(411.17,149.59){\makebox(0.00,0.00){$\hat{\alpha}_{\ell-1}$}}
\put(347.50,165.00){\makebox(0.00,0.00){$\hat{\alpha}_{\ell-2}$}}
\put(296.00,165.00){\makebox(0.00,0.00){$\hat{\alpha}_2$}}
\put(255.00,165.00){\makebox(0.00,0.00){$\hat{\alpha}_1$}}
\put(177.92,165.42){\makebox(0.00,0.00){$\alpha_\ell$}}
\put(137.92,165.42){\makebox(0.00,0.00){$\alpha_{\ell-1}$}}
\put(97.50,165.00){\makebox(0.00,0.00){$\alpha_{\ell-2}$}}
\put(45.00,165.00){\makebox(0.00,0.00){$\alpha_2$}}
\put(5.00,165.00){\makebox(0.00,0.00){$\alpha_1$}}
\put(389.00,146.33){\line(1,-1){22.67}}
\put(411.67,176.67){\line(-1,-1){23.00}}
\put(350.00,150.00){\line(1,0){30.00}}
\put(330.00,150.00){\line(1,0){10.00}}
\put(300.00,150.00){\line(1,0){10.00}}
\put(260.00,150.00){\line(1,0){30.00}}
\put(150.00,150.00){\line(1,-2){8.00}}
\put(150.00,150.00){\line(1,2){8.00}}
\put(135.33,145.00){\line(1,0){39.67}}
\put(135.33,155.00){\line(1,0){39.67}}
\put(100.00,150.00){\line(1,0){30.00}}
\put(80.00,150.00){\line(1,0){10.00}}
\put(50.00,150.00){\line(1,0){10.00}}
\put(10.00,150.00){\line(1,0){30.00}}
\put(415.00,120.00){\circle*{10.00}}
\put(415.00,180.00){\circle*{10.00}}
\put(385.00,150.00){\circle*{10.00}}
\put(345.00,150.00){\circle*{10.00}}
\put(135.00,150.00){\circle*{10.00}}
\put(175.00,150.00){\circle*{10.00}}
\put(255.00,150.00){\circle*{10.00}}
\put(295.00,150.00){\circle*{10.00}}
\put(95.00,150.00){\circle*{10.00}}
\put(45.00,150.00){\circle*{10.00}}
\put(5.00,150.00){\circle*{10.00}}
\end{picture}

\noindent The (generalized) Coxeter numbers are $h=2\ell $ , $H=2\ell +2$ in
this case. Similarly as in the previous section, we present only closed
formulae for some particles. For the first particle we find for $\ell \neq 2$
\begin{eqnarray}
\tilde{R}_{1}^{(\mathbf{C}_{\ell }^{(1)},\mathbf{D}_{\ell +1}^{(2)})}(\theta
+i\pi ,B) &=&\left\Vert h-1,\left[ h-1\right] _{2};2h-3,2h+1\right\Vert
_{\theta }\prod\limits_{k=0}^{\ell -2}\left\Vert 2k+1,2k+1\right\Vert
_{\theta }  \notag \\
&\times &\prod\limits_{k=\ell +1}^{2(\ell -1)}\left\Vert
2k-1,2k+3\right\Vert _{\theta }~,
\end{eqnarray}%
whereas for the second we obtain 
\begin{eqnarray}
\tilde{R}_{2}^{(\mathbf{C}_{\ell }^{(1)},\mathbf{D}_{\ell +1}^{(2)})}(\theta
+i\pi ,B) &=&\left\Vert 1,1;h-3,\left[ h-3\right] _{2};h-3,h-3;h-1,\left[ h-1%
\right] _{3};h+1,h+1;\right.  \notag \\
&&\left. h+1,h+5;2h-3,2h+1\right\Vert _{\theta }\prod\limits_{k=\ell
+2}^{2(\ell -1)}\left\Vert 2k-1,2k+3\right\Vert _{\theta }^{2}  \notag \\
&&\times \prod\limits_{k=0}^{\ell -4}\left\Vert 2k+3,2k+3\right\Vert
_{\theta }^{2}
\end{eqnarray}%
with $\ell \neq 2,3$. For the amplitudes related to the last two particles
we find 
\begin{eqnarray}
\tilde{R}_{\ell -1}^{(\mathbf{C}_{\ell }^{(1)},\mathbf{D}_{\ell
+1}^{(2)})}(\theta +i\pi ,B) &=&\prod\limits_{k=0}^{\ell -2}\left\Vert
2k+1,2k+1;2k+3,\left[ 2k+3\right] _{2};4k+5,4k+9\right\Vert _{\theta } 
\notag \\
&&\times \prod_{n=0}^{\ell -3}\prod\limits_{k=\ell +1-n}^{2(\ell
-n-1)}\left\Vert 2k-1,2k-1;2k-1,2k+3\right\Vert _{\theta },
\end{eqnarray}%
\begin{eqnarray}
\tilde{R}_{\ell }^{(\mathbf{C}_{\ell }^{(1)},\mathbf{D}_{\ell
+1}^{(2)})}(\theta +i\pi ,B) &=&\left\Vert 1,1_{2}\right\Vert _{\theta
}\prod\limits_{k=0}^{\ell -2}\left\Vert 2k+3,\left[ 2k+3\right]
_{3};2k+5,2k+5\right\Vert _{\theta }  \notag \\
&&\times \prod_{n=0}^{\ell -4}\prod\limits_{k=\ell -n}^{2(\ell
-n-2)}\left\Vert 2k+3,2k+3;2k+3,2k+7\right\Vert _{\theta }  \notag \\
&&\times \prod\limits_{k=1}^{\ell -2}\left\Vert
2k+5,2k+9;4k+5,4k+5\right\Vert _{\theta }.
\end{eqnarray}%
Once again we do report the remaining amplitudes as their expressions turn
out to rather lengthy.

\subsection{(F$_{4}^{(1)},$E$_{6}^{(2)}$)-affine Toda field theory}

We label the particle types according to the Dynkin diagram\medskip

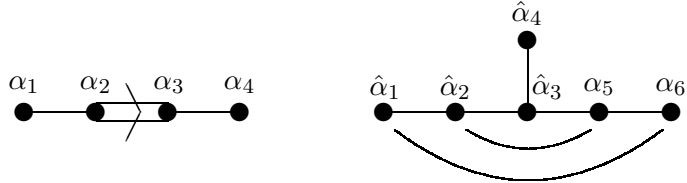
\unitlength=0.680000pt 
\begin{picture}(370.00,107.58)(-100.00,0.00)
\put(70.00,38.00){\line(-1,2){8.00}}
\put(70.00,38.00){\line(-1,-2){8.00}}
\qbezier(211.00,28.00)(285.00,-28.00)(360.00,28.00)
\qbezier(251.00,28.00)(285.00,7.01)(320.00,28.00)
\put(365.00,52.00){\makebox(0.00,0.00){$\alpha_6$}}
\put(325.00,52.00){\makebox(0.00,0.00){$\alpha_5$}}
\put(296.25,52.17){\makebox(0.00,0.00){$\hat{\alpha}_3$}}
\put(285.00,93.00){\makebox(0.00,0.00){$\hat{\alpha}_4$}}
\put(245.00,52.00){\makebox(0.00,0.00){$\hat{\alpha}_2$}}
\put(206.00,52.00){\makebox(0.00,0.00){$\hat{\alpha}_1$}}
\put(125.00,53.00){\makebox(0.00,0.00){$\alpha_4$}}
\put(85.83,53.00){\makebox(0.00,0.00){$\alpha_3$}}
\put(45.00,53.00){\makebox(0.00,0.00){$\alpha_2$}}
\put(5.00,53.00){\makebox(0.00,0.00){$\alpha_1$}}
\put(330.00,38.00){\line(1,0){30.00}}
\put(285.33,72.67){\line(0,-1){31.00}}
\put(290.00,38.00){\line(1,0){30.00}}
\put(250.00,38.00){\line(1,0){30.00}}
\put(210.00,38.00){\line(1,0){30.00}}
\put(45.33,33.00){\line(1,0){40.33}}
\put(45.33,43.00){\line(1,0){40.33}}
\put(90.00,38.00){\line(1,0){30.00}}
\put(10.00,38.00){\line(1,0){30.00}}
\put(285.00,78.00){\circle*{10.00}}
\put(245.00,38.00){\circle*{10.00}}
\put(325.00,38.00){\circle*{10.00}}
\put(365.00,38.00){\circle*{10.00}}
\put(285.00,38.00){\circle*{10.00}}
\put(45.00,38.00){\circle*{10.00}}
\put(125.00,38.00){\circle*{10.00}}
\put(5.00,38.00){\circle*{10.00}}
\put(85.00,38.00){\circle*{10.00}}
\put(205.00,38.00){\circle*{10.00}}
\end{picture}

\noindent The (generalized) Coxeter numbers are $h=12$ and $H=18$ in this
case, with $t_{1}=t_{2}=2$ and $t_{3}=t_{4}=1$. We compute 
\begin{eqnarray}
\tilde{R}_{1}(\theta +i\pi ) &=&\left\Vert
1,1_{2};3,5_{2};5,7_{3};7,9;7,9_{3};9,11_{3};9,15;11,13_{4};11,17;13,17_{2};13,23;\right.
\notag \\
&&\left. 15,21;15,21_{3};17,23;17,27;19,25;21,29_{2}\right\Vert _{\theta } \\
\tilde{R}_{2}(\theta +i\pi ) &=&\left\Vert
1,1_{2};3,3_{3};3,5_{2};5,5_{4};5,5_{3};5,9;7,7_{5};7,9_{2};7,11^{2};9,9_{5};9,13_{2};9,13^{2};\right.
\notag \\
&&11,13_{4};11,15_{3};11,15;11,19;13,17_{4};13,17^{2};13,21;15,19_{3};15,23;
\notag \\
&&\left. 15,27;17,21;17,25;17,29;19,25_{2};19,31;21,29;-19,-29\right\Vert
_{\theta } \\
\tilde{R}_{3}(\theta +i\pi ) &=&\left\Vert
1,1;3,3;3,3_{2};5,5_{3};5,7_{2};7,7_{4};7,11^{2};9,11_{4};9,13^{2};11,15_{3};11,15;11,15_{2};13,\right.
\notag \\
&&\left.
17_{3};13,21^{2};15,19;15,21_{2};15,23;17,23_{2};17,29;19,27;19,31;21,33%
\right\Vert _{\theta }~~~~ \\
\tilde{R}_{4}(\theta +i\pi ) &=&\left\Vert
1,1;3,3;5,5_{2};7,9_{2};7,11;9,13_{2};9,13;11,15_{3};13,17;13,21;15,21_{2};15,23;\right.
\notag \\
&&\left. 17,25_{2};19,31;21,33\right\Vert _{\theta }~.
\end{eqnarray}
We are not aware of any kind of solution known in the literature related to
this algebra.

\subsection{(G$_{2}^{(1)},$D$_{4}^{(3)}$)-affine Toda field theory}

We label the particle types according to the Dynkin diagram\bigskip
\smallskip

\ \ \ \unitlength=0.680000pt 
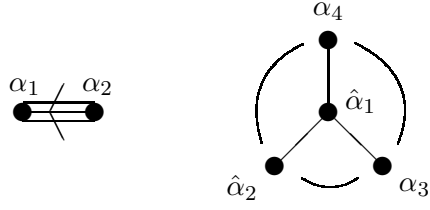
\begin{picture}(223.00,105.00)(-120.00,200.00)
\put(130.00,0.00){\line(0,1){0.00}}
\qbezier(160.00,223.00)(176.00,213.00)(191.00,223.00)
\put(223.00,218.00){\makebox(0.00,0.00){$\alpha_3$}}
\put(175.00,316.01){\makebox(0.00,0.00){$\alpha_4$}}
\put(127.00,218.00){\makebox(0.00,0.00){$\hat{\alpha}_2$}}
\put(193.00,266.00){\makebox(0.00,0.00){$\hat{\alpha}_1$}}
\put(46.67,274.17){\makebox(0.00,0.00){$\alpha_2$}}
\put(6.00,274.01){\makebox(0.00,0.00){$\alpha_1$}}
\qbezier(189.33,298.34)(227.67,279.33)(213.33,242.00)
\qbezier(160.67,297.67)(126.00,281.01)(137.67,242.33)
\put(178.67,256.33){\line(1,-1){23.33}}
\put(171.67,256.33){\line(-1,-1){23.00}}
\put(175.00,294.68){\line(0,-1){31.33}}
\put(175.00,300.00){\circle*{10.00}}
\put(205.00,230.00){\circle*{10.00}}
\put(145.00,230.00){\circle*{10.00}}
\put(175.00,260.00){\circle*{10.00}}
\put(20.00,260.00){\line(1,-2){8.00}}
\put(20.00,260.00){\line(1,2){8.00}}
\put(5.00,255.00){\line(1,0){40.00}}
\put(5.33,265.00){\line(1,0){39.67}}
\put(10.00,260.00){\line(1,0){30.00}}
\put(5.00,260.00){\circle*{10.00}}
\put(45.00,260.00){\circle*{10.00}}
\end{picture}

\noindent The (generalized) Coxeter numbers are now $h=12$ and $H=18$. In
this case we compute the integral representation $\allowbreak $ 
\begin{eqnarray}
\rho _{1}^{\mathbf{G}_{2}^{(1)}}(t) &=&\frac{16\sinh \frac{\vartheta _{h}t}{2%
}\sinh \frac{\vartheta _{H}t}{2}(\sinh \frac{t}{12}\sinh \frac{Bt}{16}-\cosh 
\frac{(B+4)t}{48}\cosh \frac{(B+4)t}{24})}{\frac{1}{2}-\cosh \frac{t}{3}%
+\cosh \frac{t}{2}}, \\
\rho _{2}^{\mathbf{G}_{2}^{(1)}}(t) &=&\frac{16\sinh \frac{\vartheta _{h}t}{2%
}\sinh \frac{3\vartheta _{H}t}{2}\left[ (2\cosh \frac{t}{6}-1)\sinh \frac{%
\vartheta _{h}t}{2}-\frac{1}{2}\cosh \frac{(B-4)t}{48}\right] }{\frac{1}{2}%
-\cosh \frac{t}{3}+\cosh \frac{t}{2}}.
\end{eqnarray}
When computing the block representation (\ref{RB}) we find complete
agreement with the solution found in \cite{Kim} shifted by $i\pi $ in the
rapidities up to some obvious typos. We therefore do not need to report it
here. The solutions differ from the ones reported in \cite{Sasaki:1993xr}.

\section{Breaking of the strong-weak duality}

The above solutions are very general and can be related easily by means of
the ambiguities (\ref{am1})-(\ref{am4}) to all other solutions which are
reported in the literature so far. Let us consider one particular ambiguity
in more detail 
\begin{equation}
\hat{R}_{i}(\theta ,B)=\tilde{R}_{i}(\theta ,B)\prod\limits_{j=1}^{\ell
}S_{ij}(\theta ,1-B/2)~.  \label{Rhat}
\end{equation}%
At first sight there seems to be nothing special about $\hat{R}_{i}(\theta
,B)$. Nonetheless, certain evident features can be seen from (\ref{Rhat}).
Our solution $\tilde{R}_{i}(\theta ,B)$ for the reflection amplitude shares
with the bulk scattering amplitude $S_{ij}(\theta ,B)$ the property of being
invariant under the strong-weak duality transformation $B\rightarrow 2-B$.
Since $S_{ij}(\theta ,1-B/2)\neq S_{ij}(\theta ,B/2)$ it is clear from (\ref%
{Rhat}) that $\hat{R}_{i}(\theta ,B)$ is not invariant under the strong-weak
transformation. As was argued in \cite{Corrigan:1994ft}, it is desirable to
construct such solutions for the reflection amplitudes, because unlike $%
\tilde{R}_{i}(\theta ,B)$ which tends to $1$ in the weak and strong
classical limit, i.e. $B\rightarrow 0,2$, we have now simply 
\begin{eqnarray}
\hat{R}_{i}(\theta ,B &=&0)=\prod\limits_{j=1}^{\ell }S_{ij}(\theta ,B=1), \\
\hat{R}_{i}(\theta ,B &=&2)=1~.
\end{eqnarray}%
This means, whilst $\tilde{R}_{i}(\theta ,B)$ reduces in the classical limit
to a theory with Neumann (free) boundary condition, the amplitude $\hat{R}%
_{i}(\theta ,B)$ tends to a theory with fixed boundary conditions for $%
B\rightarrow 0$, but for $B\rightarrow 2$ to a theory with Neumann boundary
condition. Hence the formulation (\ref{Rhat}) constitutes a simple mechanism
of breaking consistently the duality and changing from one type of boundary
conditions to another. This picture of obtaining two different classical
Lagrangians is familiar for the bulk theories of ATFT related to non-simply
laced Lie algebras and was put forward for theories with boundaries in \cite%
{Corrigan:1994ft} based on observations of the classical theory.

Let us evaluate the solution (\ref{Rhat}) in detail. From the above data and
in particular the formulae provided in the appendix, we compute for the
simply laced algebras an integral representation for $\hat{R}$ analogue to (%
\ref{Matrix}), where the corresponding kernel is 
\begin{equation}
\hat{\rho}_{i}(t)=4\frac{\sinh \frac{t(2-B)}{4h}}{\cosh \frac{t}{2}}\left[
\sinh \frac{t}{2}(1+\frac{B}{2h})\left[ K^{-1}(t/2)\right] _{ii}-2\cosh 
\frac{Bt}{4h}\sum\limits_{x\in \mathcal{\hat{X}}_{i}}\sinh \frac{xt}{2h}%
\right] ~.  \label{rhohat}
\end{equation}%
The $\mathcal{\hat{X}}_{i}$ are sets specific to the algebras and particle
types. We find (see the appendix for some details on this calculations) that 
\begin{eqnarray}
\mathcal{\hat{X}}_{i}^{\mathbf{A}_{\ell }} &=&\emptyset \qquad \text{for }%
1\leq i\leq \ell  \label{add1} \\
\mathcal{\hat{X}}_{i}^{\mathbf{D}_{\ell }} &=&\emptyset \qquad \text{for }%
i=1,\ell -1,\ell \\
\mathcal{\hat{X}}_{i}^{\mathbf{D}_{\ell }} &=&\bigcup\limits_{1\leq
k<[(2i+1)/4]}\{2i+1-4k\}\qquad \text{for }2\leq i\leq \ell -2 \\
\mathcal{\hat{X}}_{1}^{E_{6}} &=&\mathcal{\hat{X}}_{6}^{E_{6}}=\emptyset
,\quad \mathcal{\hat{X}}_{3}^{E_{6}}=\mathcal{\hat{X}}_{5}^{E_{6}}=\{5\},%
\quad \mathcal{\hat{X}}_{2}^{E_{6}}=\{1\},\quad \mathcal{\hat{X}}%
_{4}^{E_{6}}=\{3,5,7\}, \\
\mathcal{\hat{X}}_{1}^{E_{7}} &=&\{1\},\quad \mathcal{\hat{X}}%
_{2}^{E_{7}}=\{7\},\quad \mathcal{\hat{X}}_{3}^{E_{7}}=\{3,7,11\},\quad 
\mathcal{X}_{4}^{E_{7}}=\{1,5^{2},7,9^{2},11,13\}, \\
\mathcal{\hat{X}}_{5}^{E_{7}} &=&\{3,7,9,11\},\quad \mathcal{\hat{X}}%
_{6}^{E_{7}}=\{1,9\},\quad \mathcal{\hat{X}}_{7}^{E_{7}}=\emptyset , \\
\mathcal{\hat{X}}_{1}^{E_{8}} &=&\{1,13\},\quad \mathcal{\hat{X}}%
_{2}^{E_{8}}=\{7,11,15,19\},\quad \mathcal{\hat{X}}_{3}^{E_{8}}=%
\{3,7,11^{2},13,15,17,19,23\}, \\
\mathcal{\hat{X}}_{4}^{E_{8}}
&=&\{1,5^{2},7,9^{3},11^{2},13^{3},15^{2},17^{3},19^{2},21^{2},23,25\}, \\
\mathcal{\hat{X}}_{5}^{E_{8}}
&=&\{3,7^{2},9,11^{2},13,15^{2},17,19^{2},21^{2},23\}, \\
\mathcal{\hat{X}}_{6}^{E_{8}} &=&\{1,5,9,11,13,17,19,21\},\quad \mathcal{%
\hat{X}}_{7}^{E_{8}}=\{3,11,19\},\quad \mathcal{\hat{X}}_{8}^{E_{8}}=\{1\}~.
\label{add9}
\end{eqnarray}%
Up to some minor typo, the expression (\ref{rhohat}) corresponds for $%
\mathbf{A}_{\ell }$ to the formula proposed by Fateev in \cite{Fateev1},
which was obtained by changing from the block form (\ref{RB}) provided in 
\cite{Corrigan:1994ft} to an integral representation. For certain
amplitudes, namely when the Kac label $n_{i}=\psi \cdot $ $\lambda _{i}=1$,
with $\psi $ being the highest root and $\lambda _{i}$ the fundamental
weight, a conjecture was put forward in \cite{Fateev1}, which corresponds
precisely to our expression (\ref{rhohat}) when $\mathcal{\hat{X}}_{i}$ is
the empty set $\emptyset $. At present the condition for the Kac labels is
only an observation and has no deeper physical or mathematical meaning, but
probably when one computes the quantities in terms of inner products of
simple roots and weights, analogue to computations in \cite{FKS} for the
bulk S-matrix, one can provide a reasoning for it. Note that the two sets $%
\mathcal{\hat{X}}_{i}$ and $\mathcal{\tilde{X}}_{i}$ can be obtained from
each other when replacing each element $x\in \mathcal{\tilde{X}}_{i}$ by $%
(h-x)\in \mathcal{\hat{X}}_{i}$.

We may carry out the sum 
\begin{equation}
\sum\limits_{x\in \mathcal{\hat{X}}_{i}^{\mathbf{D}_{\ell }}}\sinh \frac{tx}{%
2h}=\left[ \sinh \frac{it}{2h}\sinh \frac{(i-1)t}{2h}\right] \sinh ^{-1}%
\frac{t}{h}
\end{equation}%
for $2\leq i\leq \ell -2$ and obtain the only amplitudes which were provided
in \cite{Fateev1} not satisfying the condition $n_{i}=1$. In this case we
find agreement with our solution up to a minor typo.

Alternatively, we can turn (\ref{rhohat}) into a block form formulation%
\begin{equation}
\hat{R}_{i}(\theta ,B=0)=\prod\limits_{x=1}^{h}\widehat{\left\Vert
x\right\Vert }_{\theta }^{2\mu _{ii}}\prod\limits_{x\in \mathcal{\hat{X}}%
_{i}}\overline{\left\Vert x\right\Vert }_{\theta },
\end{equation}%
where the powers $\mu _{ii}$ relate to the bulk scattering matrix as defined
in (\ref{SB}) and the blocks $\widehat{\left\Vert x\right\Vert }$, $%
\overline{\left\Vert x\right\Vert }_{\theta }$ were introduced in (\ref{bl1}%
) and (\ref{bl2}). Note that $\widehat{\left\Vert x\right\Vert }_{\theta }%
\widehat{\left\Vert x\right\Vert }_{\theta +i\pi }=\{x\}_{2\theta }$ and $%
\overline{\left\Vert x\right\Vert }_{\theta }\overline{\left\Vert
x\right\Vert }_{\theta +i\pi }=1$, such that we see that the crossing
relation (\ref{cuR}) block-wise trivially satisfied when $\hat{R}_{k}=\hat{R}%
_{\bar{k}}$.

In principle the formula (\ref{Rhat}) also holds for the non-simply laced
case and a similar reasoning as for the simply laced cases can be carried
out. However, we expect now also the occurrence of some free parameters
according to the arguments of \cite%
{Corrigan:1994ft,Bowcock:1995vp,Corrigan:1995np}. This means some
modifications are needed here. Even for special choices of the parameters
the conjecture put forward in \cite{Fateev3} does not seem to agree with (%
\ref{Rhat}). Let us briefly comment on the mechanisms, which leads to free
parameters within the bootstrap approach. We commence with the easiest model
which exhibits such features, that is the sinh-Gordon model ($A_{1}^{(1)}$%
-ATFT). Our solution for the reflection amplitude for the one particle in
the model reads in this case 
\begin{equation}
\tilde{R}(\theta ,B)=(1)_{\theta }(-B/2)_{\theta }(B/2-1)_{\theta }~.
\end{equation}%
The S-matrix is well known to be \cite{sinhG1,sinhGS2,Schroer:1976if} 
\begin{equation}
S(\theta ,B)=-(-B)_{\theta }(B-2)_{\theta }~.
\end{equation}%
We can relate our solution easily to an expression analyzed relatively
recently by Chenghlou and Corrigan \cite{sinh3} against perturbation theory.
In their notation we find 
\begin{equation}
R(\theta ,B)=\tilde{R}(\theta ,B)\frac{S(\theta ,1-E)S(\theta ,1-F)}{%
S(\theta ,B/2)S(\theta ,1+B/2)}~,  \label{RCC}
\end{equation}%
where $E$ and $F$ are free parameters. As there is no bootstrap in the
sinh-Gordon model, it is clear that every solution for $R$ multiplied by $%
S(\theta ,B^{\prime })$ constitutes also a perfectly consistent solution
from the bootstrap point of view. If then in addition the effective coupling
is taken to be in the range $0\leq B^{\prime }\leq 2$ there will be no
additional poles introduced by this multiplier, such that the bootstrap
equation (\ref{boot2}) is not coming into play. An important consequence is
that the energy of the corresponding boundary state of this solution will be
the same for all values of the free parameter $B^{\prime }$ in the stated
regime. The factors $S(\theta ,1-E)$ and $S(\theta ,1-F)$ are precisely of
this type. This argument is not yet sufficient to explain why there are
precisely two free parameters (as in principle it would allow the
introduction of an arbitrary number), but it explains when they might arise.
Similarly, we obtain a solution which was found in \cite{Baseilhac:2002kf}
for the sinh-Gordon model with dynamical boundary conditions\footnote{%
We are grateful to P.Baseihal for bringing \cite{Baseilhac:2002kf} to our
attention.}. The solution found in there relates to ours as $R(\theta ,B)=%
\tilde{R}(\theta ,B)/S(\theta ,1)$.

Let us look at a more complicated model which involves a non-trivial
bootstrap and for which we also expect this phenomenon: $%
(B_{2}^{(1)},A_{3}^{(2)})$-ATFT. In that case we can define the new
amplitudes 
\begin{equation}
R_{1}(\theta ,B,B^{\prime })\rightarrow R_{1}(\theta ,B)S_{11}(\theta
,B^{\prime })\text{ ~~and~~}R_{2}(\theta ,B,B^{\prime })\rightarrow
R_{2}(\theta ,B)S_{12}(\theta ,B^{\prime })
\end{equation}%
where the parameter $0\leq B^{\prime }\leq 2$ is kept free. Clearly there is
no problem with crossing, unitarity (\ref{cuR}) and by construction also the
boundary bootstrap equation (\ref{bootR}) is satisfied. As the amplitudes $%
S_{11}$ and $S_{12}$ introduce no new poles whose residues satisfy (\ref{res}%
), we have similarly as for sinh-Gordon a new solution whose energies of the
bound states are the same as for the original solution for all possible
values of the free parameter $B^{\prime }$. In comparison we can look at the 
$A_{2}^{(1)}$-ATFT, where such freedom does not exist. In that theory the
process $1+1$ and $2+2$ lead to new bound states, such that we can not
multiply with the corresponding S-matrices without changing the energies of
the boundary states.

We have indicated here briefly how free parameters may emerge naturally in
the bootstrap approach. A more detailed analysis of this argument we shall
present elsewhere \cite{morecomingup}.

\section{Conclusion}

In this manuscript we have provided a closed generic solution $\tilde{R}%
(\theta )$ for the boundary bootstrap equations valid for affine Toda field
theories related to all simple Lie algebras, simply laced as well as
non-simply laced. We have worked out this formula in detail for specific Lie
algebras in form of an integral representation as well as in form of blocks
of hyperbolic functions. Our solution $\tilde{R}(\theta )$ can be used as a
seed to construct (all) other solutions related to various types of boundary
conditions.

The non-uniqueness of the solution is related to the fact that one can make
use of the transformations (\ref{am1})-(\ref{am4}) and always produce new
types of solutions. The natural question which arises is: Which of these
solutions are meaningful? In the bulk theories one finds that essentially
all solutions to the bootstrap equations subjected to minimal analyticity
lead to meaningful quantum field theories. Very often there is no classical
counterpart in form of a Lagrangian known to these solutions. Even though
conceptually not needed, as an organizing principle classical Lagrangians
are very useful. In the case of boundary theories it is the different types
of boundary conditions which label the solutions (theories). In a sequence
of papers the Durham/York-group \cite%
{Corrigan:1994ft,Bowcock:1995vp,Corrigan:1995np} has investigated which type
of classical boundary terms can be used to perturb an affine Toda field
theory such that the integrability is preserved. The findings were that the
theory has to be of the form 
\begin{equation}
\mathcal{L}=\Theta (-x)\mathcal{L}_{ATFT}-\delta (x)\frac{m}{\beta ^{2}}%
\sum\limits_{i=0}^{\ell }\kappa _{i}\sqrt{n_{i}}e^{\beta \alpha _{i}\cdot
\varphi /2}  \label{bound}
\end{equation}
where the $n_{i}$ are the usual Kac labels occurring in $\mathcal{L}_{ATFT}$%
, defined through the expansion of the highest root $\psi =-\alpha
_{0}=\sum\nolimits_{i=1}^{\ell }n_{i}\alpha _{i}$ in terms of simple roots $%
\alpha _{i}$. For theories related to simply laced algebras (except
sinh-Gordon $\equiv A_{1}^{(1)}$ where the two parameters are free) the
constants $\kappa _{i}$ can be either all zero $\kappa _{i}=0$ for $\forall
i $ (Neumann boundary condition) or $\left\vert \kappa _{i}\right\vert =1$
for $\forall i$. For the non-simply laced case the $\kappa _{i}$ are fixed
depending on the algebra and there are up to two free parameters $\kappa
_{i} $ either exclusively related to the short or long roots (see appendix D
in \cite{Bowcock:1995vp} for details).

How can our solution (\ref{Matrix}) be related to the different choices of
the boundary in (\ref{bound})? Let us consider the slightly generalized
expression (\ref{Rhat}) for the $\mathbf{A}_{\ell }$-ATFT 
\begin{equation}
\hat{R}_{j}^{\pm }(\theta ,B)=\tilde{R}_{j}(\theta ,B)^{\pm
}\prod\limits_{k=1}^{\ell }S_{jk}(\theta ,1-B/2)^{\pm 1}~\text{~}.
\end{equation}%
Computing now the classical limit, we find 
\begin{eqnarray}
\lim_{B\rightarrow 0}\hat{R}_{j}^{\pm }(\theta ,B) &=&\tilde{R}_{i}(\theta
,B=0)^{\pm }\prod\limits_{k=1}^{\ell }S_{jk}(\theta ,1)^{\pm 1}  \label{l} \\
&=&\exp \left( \pm 8\int\nolimits_{0}^{\infty }\frac{dt}{t}\sinh ^{2}\frac{t%
}{2h}\sum\nolimits_{k=1}^{\ell }K_{jk}^{-1}(t)\sinh \frac{\theta t}{i\pi }%
\right)  \\
&=&\exp \left( \pm 4\int\nolimits_{0}^{\infty }\frac{dt}{t}\sinh \frac{t}{2h}%
\tanh \frac{t}{2}K_{jj}^{-1}(t/2)\sinh \frac{\theta t}{i\pi }\right)  \\
&=&-(j)_{\theta }^{\pm }(h-j)_{\theta }^{\pm } \\
&=&\frac{i\sinh \theta \mp 1/2mm_{j}}{i\sinh \theta \pm 1/2mm_{j}},
\label{l5}
\end{eqnarray}%
where we used $m_{j}=2m\sin (j\pi /h)$, with $m$ being once more an overall
mass scale. The expression (\ref{l5}) is what is predicted in this limit 
\cite{Corrigan:1994ft,Bowcock:1995vp,Corrigan:1995np}. It is then clear that
combinations of $\hat{R}_{j}^{\pm }(\theta ,B)$ for different $j$ can be
used to construct all possible fixed boundary solutions, i.e. $\hat{R}%
_{1}^{+}(\theta ,B),\hat{R}_{2}^{-}(\theta ,B),\hat{R}_{3}^{-}(\theta
,B),\ldots $ $\rightarrow $ $\{+,-,-,\ldots \}$. Similar limits can be
carried out for the other Lie algebras. For non-simply laced algebra and the
sinh-Gordon model, we gave a short argument which leads to the occurrence of
free parameter within the bootstrap approach. As many solution give the same
classical limit, it is clear that even in the simply laced cases the
classical limit\footnote{%
We do not see how this is compatible with the statement expressed in \cite%
{Gand2}, where the opposite is claimed, namely that different boundary
conditions share the same quantum reflection amplitude.} is not enough to
pin down the solutions and relate them one-to-one to one particular boundary
condition. More information can be obtained from perturbative computations,
as at order $\beta ^{2}$ already many solutions start to differ from each
other, although even at that order some distinct solutions still coincide.
Unfortunately, there are not many computations of this kind existing in the
literature to compare with.

A further way to minimize the amount of solutions which can be generated
from our generic solution $\tilde{R}$ and the ambiguities (\ref{am1})-(\ref%
{am4}) is of course to close also the second type of bootstrap equation (\ref%
{boot2}) \cite{FK3,Gand2,Riva} (see also section 4.6.1 for an example
related to non-simply laced Lie algebras) and eliminate those solutions
which do not allow for such a closure. A systematic study of this kind has
not been carried out so far and we share the pessimistic viewpoint expressed
in \cite{Gand2} concerning such an undertaking. Whereas it appears possible
to show that some solutions do indeed close, it seems difficult to develop a
systematic scheme which selects solution which do not close.\ Possibly when
developing a formulation in terms of Coxeter geometry similarly as in the
bulk \cite{FKS}, this can be understood better.

More in the spirit of exactly solvable models are considerations carried out
in \cite{Fat2,Fat3}, where the scaling functions (free energies) have been
computed in two alternative ways. On one hand one can compute it by means of
the thermodynamic Bethe ansatz and on the other by a semi-classical
perturbation around the conformal field theory. Since in the former the
boundary reflection amplitude enters as an input and in the latter the
explicit boundary conditions one may compare the outcome and therefore
indirectly relate solutions of the boundary bootstrap equations and
classical boundary conditions. We leave this analysis for future
investigations \cite{comingup}.

\medskip

\textbf{Acknowledgments: }We are grateful to the Deutsche
Forschungsgemeinschaft (Sfb288), for financial support. This work is
supported by the EU network \textquotedblleft EUCLID, \emph{Integrable
models and applications: from strings to condensed matter}\textquotedblright
, HPRN-CT-2002-00325. We \ are grateful to G.W. Delius and M. Stanishkov for
discussions.

\appendix

\section{Appendix: The inverse of K and the evaluation of $\protect\rho $}

The central object which enters into the computation of the bulk scattering
amplitude (\ref{white}) as well as into the reflection amplitude (\ref%
{Matrix}) is the inverse of the q-deformed Cartan matrix (\ref{Kq}). We
demonstrate that the entries of this matrix can be written in a closed form
and furthermore that the sums over rows or columns can be carried out
explicitly. Also this closed form can be used when they enter the expression
for the kernel in (\ref{rho}) and (\ref{rhohat}).

Let us comment on the evaluation for this object in this appendix. We only
present the formulae for the simply laced algebras $\mathbf{g}$ and first
determine the determinant of $K(t)$. For simply laced algebras we know the
eigenvalues of the incidence matrix $I$ of $\mathbf{g}$ to be $%
I_{ij}y_{j}(n)=2\cos (\pi s_{n}/h)y_{i}(n)$, where the $s_{n}$ are the
exponents of $\mathbf{g}$. Therefore we conclude directly 
\begin{equation}
K_{ij}^{\mathbf{g}}(t)y_{j}(n)=4\cosh \left[ (t+i\pi s_{n})/2h\right] \cosh %
\left[ (t-i\pi s_{n})/2h\right] y_{i}(n)=\lambda _{n}^{\mathbf{g}}y_{i}(n)~.
\label{la}
\end{equation}
Appealing to the well-known relation between the eigenvalues of a matrix and
its determinant we obtain 
\begin{equation}
\det K^{\mathbf{g}}(t)=\prod\limits_{n=1}^{\ell }\lambda _{n}^{\mathbf{g}%
}=\prod\limits_{n=1}^{\ell }4\cosh \left[ (t+i\pi s_{n})/2h\right] \cosh %
\left[ (t-i\pi s_{n})/2h\right] ~.  \label{tK}
\end{equation}
Having in mind to compute the inverse of $K(t)$ we also need to determine
its cofactors. It turns out that the sub-matrix resulting from the
elimination of the $i$-th row and the $j$-th column always decomposes into
some matrices which can be identified as a deformed Cartan matrix of some
new algebras $\mathbf{\tilde{g}}_{i}$ and $\mathbf{\tilde{g}}_{j}$%
\begin{equation}
K^{\mathbf{g/\tilde{g}}}(t)\rightarrow K^{\mathbf{g}/\mathbf{\tilde{g}}%
_{i}}(t)\oplus K^{\mathbf{g}/\mathbf{\tilde{g}}_{j}}(t)~.  \label{dsum}
\end{equation}
Here we introduced the matrices 
\begin{equation}
K^{\mathbf{g/\tilde{g}}}(t):=2\cosh (t/h)-I^{\mathbf{\tilde{g}}}~.
\end{equation}
Hence $K^{\mathbf{g/\tilde{g}}}(t)$ differs from $K^{\mathbf{\tilde{g}}}(t)$
in the sense that the Coxeter number $h$ appearing in its diagonal belongs
to $\mathbf{g}$ rather than $\mathbf{\tilde{g}}$. The same argument which
lead to (\ref{la}) then gives the eigenvalues of $K^{\mathbf{g/\tilde{g}}%
}(t) $%
\begin{equation}
\lambda _{n}^{\mathbf{g/\tilde{g}}}=4\cosh (t/2h+i\pi \tilde{s}_{n}/2\tilde{h%
})\cosh (t/2h+i\pi \tilde{s}_{n}/2\tilde{h})~.
\end{equation}
Therefore we obtain the inverse of the (doubly) q-deformed Cartan matrix 
\begin{equation}
\left[ K^{\mathbf{g}}(t)^{-1}\right] _{ij}=\frac{\det K^{\mathbf{g}/\mathbf{%
\tilde{g}}_{i}}(t)\det K^{\mathbf{g}/\mathbf{\tilde{g}}_{j}}(t)}{\det K^{%
\mathbf{g}}(t)}=\frac{\prod\nolimits_{n=1}^{\ell _{i}}\lambda _{n}^{\mathbf{g%
}/\mathbf{\tilde{g}}_{i}}\prod\nolimits_{n=1}^{\ell _{j}}\lambda _{n}^{%
\mathbf{g}/\mathbf{\tilde{g}}_{j}}}{\prod\nolimits_{n=1}^{\ell }\lambda
_{n}^{\mathbf{g}}}~.  \label{KIeigen}
\end{equation}
What remains to be specified is the precise decomposition (\ref{dsum}). We
shall demonstrate this in detail. For this we need to specialize the formula
(\ref{KIeigen}) for some concrete algebras. As (\ref{KIeigen}) consists of
products it is not very suitable in that form and we therefore also present
some alternative method which turns the products into sums. We also need to
compute the sum over some rows or columns of $K(t)^{-1}$ and then we
evaluate the sums in (\ref{rho}).

\subsection{$\mathbf{A}_{\ell }$}

Taking $\mathbf{g}$\textbf{\ }to be $\mathbf{A}_{\ell }$ in (\ref{KIeigen})
it is easy to convince oneself that 
\begin{eqnarray}
\left[ K^{\mathbf{A}_{\ell }}(t)\right] _{ij}^{-1} &=&\left[ K^{\mathbf{A}%
_{\ell }}(t)\right] _{ji}^{-1}=\frac{\det K^{\mathbf{A}_{\ell }/\mathbf{A}%
_{i-1}}\det K^{\mathbf{A}_{\ell }/\mathbf{A}_{n-j}}}{\det K^{\mathbf{A}%
_{\ell }}}\qquad \text{for~~}i\leq j  \label{e1} \\
&=&\frac{\prod\nolimits_{n=1}^{i-1}\lambda _{n}^{\mathbf{A}_{\ell }/\mathbf{A%
}_{i-1}}\prod\nolimits_{n=1}^{\ell -j}\lambda _{n}^{\mathbf{A}_{\ell }/%
\mathbf{A}_{\ell -j}}}{\prod\nolimits_{n=1}^{\ell }\lambda _{n}^{\mathbf{A}%
_{\ell }}}~.  \label{e2}
\end{eqnarray}
Having in mind to sum over some rows and columns of $K(t)^{-1}$, we present
a different method to compute (\ref{e1}). For this we develop the
determinant of $K^{\mathbf{A}_{\ell }/\mathbf{A}_{n}}$ with respect to the
first row or column 
\begin{equation}
\det K^{\mathbf{A}_{\ell }/\mathbf{A}_{n}}=\det K^{\mathbf{A}_{\ell }/%
\mathbf{A}_{1}}\det K^{\mathbf{A}_{\ell }/\mathbf{A}_{n-1}}-\det K^{\mathbf{A%
}_{\ell }/\mathbf{A}_{n-2}}~.  \label{rec}
\end{equation}
Understanding that $\det K^{\mathbf{A}_{\ell }/\mathbf{A}_{0}}=1$ and $\det
K^{\mathbf{A}_{\ell }/\mathbf{A}_{n}}=0$ for $n<0$, we can view (\ref{rec})
as a recursive equation for $\det K^{\mathbf{A}_{\ell }/\mathbf{A}_{n}}$ in
terms of $\det K^{\mathbf{A}_{\ell }/\mathbf{A}_{1}}=K^{\mathbf{A}_{\ell }/%
\mathbf{A}_{1}}$, which we can leave completely arbitrary at this point. We
note that the equation (\ref{rec}) is the recursive equation for the
Chebychev polynomials of the second kind $U_{n}(x)$, such that 
\begin{equation}
\det K^{\mathbf{A}_{\ell }/\mathbf{A}_{n}}=\sum\limits_{k=0}^{[n/2]}(-1)^{k}%
\binom{n-k}{k}\left( K^{\mathbf{A}_{\ell }/\mathbf{A}_{1}}\right)
^{n-2k}=U_{n}\left( K^{\mathbf{A}_{\ell }/\mathbf{A}_{1}}/2\right) ~,
\label{aln}
\end{equation}
where $\left[ x\right] $ denotes the integer part of $x$. We also need below 
\begin{equation}
\sum\limits_{n=0}^{p}\det K^{\mathbf{A}_{\ell }/\mathbf{A}%
_{n}}=\sum\limits_{n=0}^{p}\sum\limits_{k=0}^{[n/2]}(-1)^{k}\binom{p-n+k}{k}%
\left( K^{\mathbf{A}_{\ell }/\mathbf{A}_{1}}\right) ^{p-n}.  \label{ss}
\end{equation}
Let us now fix $K^{\mathbf{A}_{\ell }/\mathbf{A}_{1}}=q+q^{-1}$. Then we
obtain from (\ref{aln}) 
\begin{equation}
\det K^{\mathbf{A}_{\ell }/\mathbf{A}_{n}}=U_{n}\left[ (q+q^{-1})/2\right] =%
\left[ 1+n\right] _{q},  \label{ff1}
\end{equation}
such that (\ref{e1}) yields 
\begin{equation}
\left[ K^{\mathbf{A}_{\ell }}(t)\right] _{ij}^{-1}=\left[ K^{\mathbf{A}%
_{\ell }}(t)\right] _{ji}^{-1}=\frac{\left[ i\right] _{q}\left[ h-j\right]
_{q}}{\left[ h\right] _{q}}\qquad \text{for~~}i\leq j~.  \label{KIA}
\end{equation}
The same specialization reduces (\ref{ss}) to 
\begin{equation}
\sum\limits_{n=0}^{p-1}\det K^{\mathbf{A}_{\ell }/\mathbf{A}_{n}}=\left[
(p+1)/2\right] _{q}\left[ p\right] _{q^{1/2}}~.  \label{f2}
\end{equation}
We can now also carry out the sum over the $i$-th row or column. From (\ref%
{e1}), (\ref{ff1}), (\ref{KIA}) and (\ref{f2}) follows 
\begin{eqnarray}
\sum\limits_{j=1}^{\ell }[K^{\mathbf{A}_{\ell }}(t)]_{ij}^{-1} &=&\frac{1}{%
\det K^{\mathbf{A}_{\ell }}}\left[ \det K^{\mathbf{A}_{\ell }/\mathbf{A}%
_{i-1}}\sum\limits_{j=0}^{\ell -i}\det K^{\mathbf{A}_{\ell }/\mathbf{A}%
_{j}}+\det K^{\mathbf{A}_{\ell }/\mathbf{A}_{\ell
-i}}\sum\limits_{j=0}^{i-2}\det K^{\mathbf{A}_{\ell }/\mathbf{A}_{j}}\right]
\notag \\
&=&\frac{1}{\left[ h\right] _{q}}\left( \left[ i\right] _{q}\left[ (h+1-i)/2%
\right] _{q}\left[ h-i\right] _{q^{1/2}}+\left[ h-i\right] _{q}\left[ i/2%
\right] _{q}\left[ i-1\right] _{q^{1/2}}\right) ,  \notag \\
&=&\frac{1}{2\cosh t/2}\left[ h-i\right] _{q^{1/2}}\left[ i\right]
_{q^{1/2}}~,  \label{s1} \\
&=&\frac{\tanh (t/2)}{2\sinh (t/2h)}(K^{\mathbf{A}_{\ell }}(t/2))_{ii}^{-1}~.
\notag
\end{eqnarray}
To be able to compute (\ref{rho}) in more detail we derive from the above
relations 
\begin{equation}
\lbrack K^{\mathbf{A}_{\ell }}(t)]_{ij}^{-1}\,\chi _{j}^{kp}(t)[K^{\mathbf{A}%
_{\ell }}(t/2)]_{kp}^{-1}=\frac{\cosh (t/2h)\sinh [(1-h)t/2h]}{2\cosh
t/2\sinh (t/h)}(K^{\mathbf{A}_{\ell }}(t/2))_{ii}^{-1}~.  \label{s2}
\end{equation}
It is the non obvious feature that the sum in (\ref{s1}) as well as the
expressions in (\ref{s1}) are both proportional to $(K(t/2))_{ii}^{-1}~$%
which allows for the computation of (\ref{rhohat}). For the other algebras
there are additional terms appearing as indicated in (\ref{add1})-(\ref{add9}%
). We proceed similarly for them.

\subsection{$\mathbf{D}_{\ell }$}

As in the previous section we find also a recursive relation for this case
by expanding the determinant with respect to the first (last) row or column 
\begin{eqnarray}
\det K^{\mathbf{D}_{\ell }/\mathbf{D}_{n}} &=&K^{\mathbf{D}_{\ell }/\mathbf{A%
}_{1}}\det K^{\mathbf{D}_{\ell }/\mathbf{D}_{n-1}}-\det K^{\mathbf{D}_{\ell
}/\mathbf{D}_{n-2}}, \\
&=&K^{\mathbf{D}_{\ell }/\mathbf{A}_{1}}\left[ \det K^{\mathbf{D}_{\ell }/%
\mathbf{A}_{n-1}}-\det K^{\mathbf{D}_{\ell }/\mathbf{A}_{n-3}}\right] ~. 
\notag
\end{eqnarray}%
From the second equality we observe that we can employ the results of the
previous section and express the determinant in terms of sub-determinants
related to $A_{n}$ algebras. In this way we compute the following expansion
in terms of $K^{\mathbf{D}_{\ell }/\mathbf{A}_{1}},$%
\begin{equation}
\det K^{\mathbf{D}_{\ell }/\mathbf{D}_{n}}=\sum\limits_{k=0}^{[n/2]}(-1)^{k}%
\frac{n-1}{n-k-1}\left( 
\begin{array}{c}
n-k-1 \\ 
k%
\end{array}%
\right) \left( K^{\mathbf{D}_{\ell }/\mathbf{A}_{1}}\right) ^{n-2k}.
\end{equation}%
When we fix $K^{\mathbf{D}_{\ell }/\mathbf{A}_{1}}=q+q^{-1}=2\cosh t/h$,
this equation becomes 
\begin{equation}
\det K^{\mathbf{D}_{\ell }/\mathbf{D}_{n}}=\left[ 2\right] _{q}(\left[ \ell
-n\right] _{q}-\left[ \ell -n-2\right] _{q})=4\cosh \frac{t(\ell -n-1)}{h}%
\cosh \frac{t}{h}.
\end{equation}%
The object which enters the general formulae for the reflection amplitudes
is the inverse of the $q$-deformed Cartan matrix. With the help of the
previous equalities, we can compute the cofactors and find 
\begin{eqnarray}
\left[ K^{\mathbf{D}_{\ell }}(t)\right] _{ij}^{-1} &=&\left[ K^{\mathbf{D}%
_{\ell }}(t)\right] _{ji}^{-1}=\frac{\sinh it/h\cosh (\ell -j-1)t/h}{\sinh
t/h\cosh t/2}\quad \text{for~}1\leq i\leq j\leq \ell -2,\quad ~~
\label{Dij1} \\
\left[ K^{\mathbf{D}_{\ell }}(t)\right] _{ip}^{-1} &=&\left[ K^{\mathbf{D}%
_{\ell }}(t)\right] _{pi}^{-1}=\frac{\sinh it/h}{2\sinh t/h\cosh t/2}\qquad
\qquad \qquad \text{for~\quad }p=\ell ,\ell -1,  \label{Dil}
\end{eqnarray}%
together with 
\begin{eqnarray}
\left[ K^{\mathbf{D}_{\ell }}(t)\right] _{\ell \ell -1}^{-1} &=&\left[ K^{%
\mathbf{D}_{\ell }}(t)\right] _{\ell -1\ell }^{-1}=\frac{\sinh (\ell -2)t/h}{%
2\sinh 2t/h\cosh t/2},  \label{Dll1} \\
\left[ K^{\mathbf{D}_{\ell }}(t)\right] _{\ell \ell }^{-1} &=&\left[ K^{%
\mathbf{D}_{\ell }}(t)\right] _{\ell -1\ell -1}^{-1}=\frac{\sinh \ell t/h}{%
2\sinh 2t/h\cosh t/2}.  \label{Dll}
\end{eqnarray}%
Taking now the sums over a row or a column gives 
\begin{equation}
\sum\limits_{j=1}^{\ell }\left[ K^{\mathbf{D}_{\ell }}(t)\right] _{ij}^{-1}=%
\frac{\sinh \frac{it}{h}}{2\cosh \frac{t}{2}\sinh \frac{t}{h}}+\frac{\sinh 
\frac{it}{2h}\sinh \frac{(h-i)t}{2h}}{\tanh \frac{t}{2h}\cosh \frac{t}{2}%
\sinh \frac{t}{h}}\quad \text{for}~1\leq i\leq j\leq \ell -2,~~  \label{54}
\end{equation}%
and 
\begin{equation}
\sum\limits_{j=1}^{\ell }\left[ K^{\mathbf{D}_{\ell }}(t)\right] _{j\ell
}^{-1}=\sum\limits_{j=1}^{\ell }\left[ K^{\mathbf{D}_{\ell }}(t)\right]
_{j\ell -1}^{-1}=\frac{\sinh \frac{\ell t}{2h}\sinh \frac{t}{4}}{2\cosh 
\frac{t}{2}\sinh \frac{t}{h}\sinh \frac{t}{2h}}.  \label{dsuml}
\end{equation}%
Combining now (\ref{Dij1})-(\ref{Dil}) and (\ref{54})-(\ref{dsuml}), we
obtain 
\begin{equation}
\frac{\sum_{j=1}^{\ell }\left[ K^{\mathbf{D}_{\ell }}(t)\right] _{ij}^{-1}}{%
\left[ K^{\mathbf{D}_{\ell }}(t/2)\right] _{ii}^{-1}}=\frac{\tanh t/2}{%
2\sinh t/2h},\quad \quad \text{for\quad }i=1,\ell -1,\ell ,  \label{r1d}
\end{equation}%
and 
\begin{equation}
\frac{\sum_{j=1}^{\ell }\left[ K^{\mathbf{D}_{\ell }}(t)\right] _{ij}^{-1}}{%
\left[ K^{\mathbf{D}_{\ell }}(t/2)\right] _{ii}^{-1}}=\frac{\left( \cosh 
\frac{it}{2h}\sinh \frac{t}{2h}+\sinh \frac{(h-i)t}{2h}\cosh \frac{t}{2h}%
\right) \cosh \frac{t}{4}}{\cosh \frac{t}{2}\sinh \frac{t}{h}\cosh \frac{%
(h-2i)t}{4h}}.  \label{r2d}
\end{equation}%
for the remaining values $2\leq i\leq \ell -2$. Finally, we may compute the
quantity, which enters directly our expressions for $R_{i}(\theta )$. We
find the following closed formulae for ~$1\leq i\leq \ell -2$ 
\begin{equation}
\sum_{j,k,p=1}^{\ell }\left[ K^{\mathbf{D}_{\ell }}(t)\right] _{ij}^{-1}\chi
_{j}^{kp}\left[ K^{\mathbf{D}_{\ell }}(t/2)\right] _{kp}^{-1}=\frac{4\cosh 
\frac{(h-2)t}{4h}\sinh \frac{t}{4}\sinh \frac{(i-h)t}{2h}\sinh \frac{it}{2h}%
}{\sinh t\sinh ^{2}\frac{t}{2h}},~\ 
\end{equation}%
and 
\begin{equation}
\sum_{j,k,p=1}^{\ell }\left[ K^{\mathbf{D}_{\ell }}(t)\right] _{ij}^{-1}\chi
_{j}^{kp}\left[ K^{\mathbf{D}_{\ell }}(t/2)\right] _{kp}^{-1}=\frac{2\sinh 
\frac{(1-h)t}{2h}\sinh \frac{(h+1-2\left[ \ell /2\right] )t}{2h}\sinh \frac{t%
\left[ \ell /2\right] }{h}}{\sinh t\sinh \frac{t}{h}\sinh \frac{t}{2h}},
\end{equation}%
for $i=\ell -1,\ell $. Having these formulae at hand we can easily obtain
the kernels (\ref{rho1}) and (\ref{rho2}). In this case we found that the
proportionality to $(K(t/2))_{ii}^{-1}$, observed in the previous section,
no longer holds and therefore the formulae are more lengthy when $i\neq
1,\ell -1,\ell $

\subsection{$\mathbf{E}_{6}$}

Developing the determinant gives again some recursive equation 
\begin{eqnarray}
\det K^{\mathbf{E}_{6}} &=&\left[ K^{\mathbf{E}_{6}/\mathbf{A}_{1}}\right]
^{2}\left[ \det K^{\mathbf{E}_{6}/\mathbf{A}_{4}}-\det K^{\mathbf{E}_{6}/%
\mathbf{A}_{2}}\right] -\det K^{\mathbf{E}_{6}/\mathbf{A}_{4}}  \notag \\
&=&K^{\mathbf{E}_{6}/\mathbf{A}_{1}}\det K^{\mathbf{E}_{6}/\mathbf{D}%
_{5}}-\det K^{\mathbf{E}_{6}/\mathbf{D}_{4}}.
\end{eqnarray}
We note that we can use once more the results of the previous sections.
Specifying $K^{\mathbf{E}_{6}/\mathbf{A}_{1}}=2\cosh t/12$ we compute the
cofactors and obtain the inverse of the q-deformed Cartan matrix for the
simply-laced algebra $E_{6}$ 
\begin{equation}
\left[ K^{\mathbf{E}_{6}}(t)\right] ^{-1}=\frac{1}{\mathbf{E}_{6}}\left( 
\begin{array}{cccccc}
\mathbf{D}_{5} & \mathbf{A}_{2} & \mathbf{A}_{4} & \mathbf{A}_{1}\mathbf{A}%
_{2} & \mathbf{A}_{1}^{2} & \mathbf{A}_{1} \\ 
\mathbf{A}_{2} & \mathbf{A}_{5} & \mathbf{A}_{1}\mathbf{A}_{2} & \mathbf{A}%
_{2}^{2} & \mathbf{A}_{1}\mathbf{A}_{2} & \mathbf{A}_{2} \\ 
\mathbf{A}_{4} & \mathbf{A}_{1}\mathbf{A}_{2} & \mathbf{A}_{1}\mathbf{A}_{4}
& \mathbf{A}_{1}^{2}\mathbf{A}_{2} & \mathbf{A}_{1}^{3} & \mathbf{A}_{1}^{2}
\\ 
\mathbf{A}_{1}\mathbf{A}_{2} & \mathbf{A}_{2}^{2} & \mathbf{A}_{1}^{2}%
\mathbf{A}_{2} & \mathbf{A}_{2}^{2}\mathbf{A}_{1} & \mathbf{A}_{1}^{2}%
\mathbf{A}_{2} & \mathbf{A}_{1}\mathbf{A}_{2} \\ 
\mathbf{A}_{1}^{2} & \mathbf{A}_{1}\mathbf{A}_{2} & \mathbf{A}_{1}^{3} & 
\mathbf{A}_{1}^{2}\mathbf{A}_{2} & \mathbf{A}_{1}\mathbf{A}_{4} & \mathbf{A}%
_{4} \\ 
\mathbf{A}_{1} & \mathbf{A}_{2} & \mathbf{A}_{1}^{2} & \mathbf{A}_{1}\mathbf{%
A}_{2} & \mathbf{A}_{4} & \mathbf{D}_{5}%
\end{array}
\right)
\end{equation}
where we understand the entries of this matrix as $g\equiv \det K^{\mathbf{E}%
_{6}/\mathbf{g}}$. Taking now the sum of particular rows and column gives 
\begin{eqnarray}
\sum\limits_{j=1}^{6}\left[ K^{\mathbf{E}_{6}}(t)\right] _{1j}^{-1}
&=&\sum\limits_{j=1}^{6}\left[ K^{\mathbf{E}_{6}}(t)\right] _{6j}^{-1}=\frac{%
2+2\sum_{k=1}^{3}\cosh \frac{kt}{12}}{2\cosh \frac{t}{3}-1}, \\
\sum\limits_{j=1}^{6}\left[ K^{\mathbf{E}_{6}}(t)\right] _{3j}^{-1}
&=&\sum\limits_{j=1}^{6}\left[ K^{\mathbf{E}_{6}}(t)\right] _{5j}^{-1}=\frac{%
3+2\sum_{k=1}^{3}(4-k)\cosh \frac{kt}{12}}{2\cosh \frac{t}{3}-1}, \\
\sum\limits_{j=1}^{6}\left[ K^{\mathbf{E}_{6}}(t)\right] _{2j}^{-1} &=&\frac{%
3+2\cosh \frac{t}{12}+2\sum_{k=1}^{3}\cosh \frac{kt}{12}}{2\cosh \frac{t}{3}%
-1}, \\
\sum\limits_{j=1}^{6}\left[ K^{\mathbf{E}_{6}}(t)\right] _{4j}^{-1} &=&\frac{%
5+8\cosh \frac{t}{12}+6\cosh \frac{t}{6}+2\cosh \frac{t}{4}}{2\cosh \frac{t}{%
3}-1}.
\end{eqnarray}
From this we compute

\begin{eqnarray*}
\sum_{j,k,p=1}^{6}\left[ K^{\mathbf{E}_{6}}(t)\right] _{1j}^{-1}\chi
_{j}^{kp}\left[ K^{\mathbf{E}_{6}}(\frac{t}{2})\right] _{kp}^{-1} &=&-\frac{%
1+2\sum_{k=1}^{5}\cosh \frac{kt}{12}}{\cosh \frac{t}{2}}\left[ K^{\mathbf{E}%
_{6}}(\frac{t}{2})\right] _{11}^{-1}, \\
\sum_{j,k,p=1}^{6}\left[ K^{\mathbf{E}_{6}}(t)\right] _{2j}^{-1}\chi
_{j}^{kp}\left[ K^{\mathbf{E}_{6}}(\frac{t}{2})\right] _{kp}^{-1} &=&\frac{%
2-4\left[ \sum\limits_{k=1}^{2}\cosh \frac{kt}{12}+\cosh \frac{t}{3}\right] 
}{\left( 1-2\cosh \frac{t}{3}\right) \left( 1-2\cosh \frac{t}{12}\right) }%
\left[ K^{\mathbf{E}_{6}}(\frac{t}{2})\right] _{22}^{-1},~~~~~ \\
\sum_{j,k,p=1}^{6}\left[ K^{\mathbf{E}_{6}}(t)\right] _{3j}^{-1}\chi
_{j}^{kp}\left[ K^{\mathbf{E}_{6}}(\frac{t}{2})\right] _{kp}^{-1} &=&\frac{%
4\sum_{k=2}^{3}\cosh \frac{kt}{12}}{1-2\cosh \frac{t}{3}}\left[ K^{\mathbf{E}%
_{6}}(\frac{t}{2})\right] _{33}^{-1}, \\
\sum_{j,k,p=1}^{6}\left[ K^{\mathbf{E}_{6}}(t)\right] _{4j}^{-1}\chi
_{j}^{kp}\left[ K^{\mathbf{E}_{6}}(\frac{t}{2})\right] _{kp}^{-1} &=&\frac{%
2\left( 1+2\cosh \frac{t}{4}\right) }{1-2\cosh \frac{t}{3}}\left[ K^{\mathbf{%
E}_{6}}(\frac{t}{2})\right] _{44}^{-1}.
\end{eqnarray*}
Taking the first subscript to be $5$ or $6$ equals the expressions for
taking them to be $3$ or $1$, respectively. This is sufficient to compute
the expressions for $R$.

\subsection{$\mathbf{E}_{7}$}

Developing the determinant gives now the recursive equations 
\begin{eqnarray}
\det K^{\mathbf{E}_{7}} &=&\left[ K^{\mathbf{E}_{7}/\mathbf{A}_{1}}\right]
^{2}\left[ \det K^{\mathbf{E}_{7}/\mathbf{A}_{5}}-\det K^{\mathbf{E}_{7}/%
\mathbf{A}_{3}}\right] -\det K^{\mathbf{E}_{7}/\mathbf{A}_{5}}  \notag \\
&=&K^{\mathbf{E}_{7}/\mathbf{A}_{1}}\det K^{\mathbf{E}_{7}/\mathbf{D}%
_{6}}-\det K^{\mathbf{E}_{7}/\mathbf{A}_{5}},
\end{eqnarray}
which can be evaluated again from the quantities already computed in the
previous sections. Specializing $K^{\mathbf{E}_{7}/\mathbf{A}_{1}}=2\cosh
t/18$ we compute 
\begin{equation}
\left[ K^{\mathbf{E}_{7}}(t)\right] ^{-1}=\frac{1}{\mathbf{E}_{7}}\left( 
\begin{array}{ccccccc}
\mathbf{D}_{6} & \mathbf{A}_{3} & \mathbf{A}_{5} & \mathbf{A}_{1}\mathbf{A}%
_{3} & \mathbf{A}_{1}\mathbf{A}_{2} & \mathbf{A}_{1}^{2} & \mathbf{A}_{1} \\ 
\mathbf{A}_{3} & \mathbf{A}_{6} & \mathbf{A}_{1}\mathbf{A}_{3} & \mathbf{A}%
_{2}\mathbf{A}_{3} & \mathbf{A}_{2}^{2} & \mathbf{A}_{1}\mathbf{A}_{2} & 
\mathbf{A}_{2} \\ 
\mathbf{A}_{5} & \mathbf{A}_{1}\mathbf{A}_{3} & \mathbf{A}_{1}\mathbf{A}_{5}
& \mathbf{A}_{1}^{2}\mathbf{A}_{3} & \mathbf{A}_{1}^{2}\mathbf{A}_{2} & 
\mathbf{A}_{1}^{3} & \mathbf{A}_{1}^{2} \\ 
\mathbf{A}_{1}\mathbf{A}_{3} & \mathbf{A}_{2}\mathbf{A}_{3} & \mathbf{A}%
_{1}^{2}\mathbf{A}_{3} & \mathbf{A}_{1}\mathbf{A}_{2}\mathbf{A}_{3} & 
\mathbf{A}_{2}^{2}\mathbf{A}_{1} & \mathbf{A}_{1}^{2}\mathbf{A}_{2} & 
\mathbf{A}_{1}\mathbf{A}_{2} \\ 
\mathbf{A}_{1}\mathbf{A}_{2} & \mathbf{A}_{2}^{2} & \mathbf{A}_{1}^{2}%
\mathbf{A}_{2} & \mathbf{A}_{2}^{2}\mathbf{A}_{1} & \mathbf{A}_{2}\mathbf{A}%
_{4} & \mathbf{A}_{1}\mathbf{A}_{4} & \mathbf{A}_{4} \\ 
\mathbf{A}_{1}^{2} & \mathbf{A}_{1}\mathbf{A}_{2} & \mathbf{A}_{1}^{3} & 
\mathbf{A}_{1}^{2}\mathbf{A}_{2} & \mathbf{A}_{1}\mathbf{A}_{4} & \mathbf{A}%
_{1}\mathbf{D}_{5} & \mathbf{D}_{5} \\ 
\mathbf{A}_{1} & \mathbf{A}_{2} & \mathbf{A}_{1}^{2} & \mathbf{A}_{1}\mathbf{%
A}_{2} & \mathbf{A}_{4} & \mathbf{D}_{5} & \mathbf{E}_{6}%
\end{array}
\right)
\end{equation}
We abbreviated here $g\equiv \det K^{\mathbf{E}_{7}/\mathbf{g}}$. The sum of
particular rows and column gives 
\begin{eqnarray}
\sum\limits_{j=1}^{7}\left[ K^{\mathbf{E}_{7}}(t)\right] _{1j}^{-1} &=&\frac{%
3+4\sum_{k=1}^{2}\cosh \frac{kt}{18}+2\sum_{k=3}^{5}\cosh \frac{kt}{18}}{%
2\cosh \frac{t}{3}-1}, \\
\sum\limits_{j=1}^{7}\left[ K^{\mathbf{E}_{7}}(t)\right] _{2j}^{-1} &=&\frac{%
\left[ -1+2\cosh \frac{t}{18}\right] \left[ 1+2\sum_{k=1}^{2}\cosh \frac{kt}{%
18}\right] ^{2}-\frac{1}{2}\cosh ^{-1}\frac{t}{18}}{2\cosh \frac{t}{3}-1}, \\
\sum\limits_{j=1}^{7}\left[ K^{\mathbf{E}_{7}}(t)\right] _{3j}^{-1} &=&\frac{%
5\left[ 1+2\cosh \frac{t}{18}\right] +6\sum_{k=2}^{3}\cosh \frac{kt}{18}+2%
\left[ 2\cosh \frac{2t}{9}+\cosh \frac{5t}{18}\right] }{2\cosh \frac{t}{3}-1}%
,
\end{eqnarray}
\begin{eqnarray}
\sum\limits_{j=1}^{7}\left[ K^{\mathbf{E}_{7}}(t)\right] _{4j}^{-1}
&=&\sum\limits_{j=1}^{7}\left[ K^{\mathbf{E}_{7}}(t)\right] _{3j}^{-1}+\frac{%
3+6\cosh \frac{t}{9}+2\left[ \sum_{k=3}^{4}\cosh \frac{kt}{18}+\cosh \frac{t%
}{18}\right] }{2\cosh \frac{t}{3}-1}, \\
\sum\limits_{j=1}^{7}\left[ K^{\mathbf{E}_{7}}(t)\right] _{5j}^{-1}
&=&\sum\limits_{j=1}^{7}\left[ K^{\mathbf{E}_{7}}(t)\right] _{3j}^{-1}+\frac{%
2\sum_{k=2}^{3}\cosh \frac{kt}{18}+\frac{1}{2}\cosh ^{-1}\frac{t}{18}}{%
2\cosh \frac{t}{3}-1}, \\
\sum\limits_{j=1}^{7}\left[ K^{\mathbf{E}_{7}}(t)\right] _{6j}^{-1}
&=&\sum\limits_{j=1}^{7}\left[ K^{\mathbf{E}_{7}}(t)\right] _{2j}^{-1}+\frac{%
1-2\left[ \cosh \frac{t}{18}+\cosh \frac{2t}{9}\right] +\frac{1}{2}\cosh
^{-1}\frac{t}{18}}{2\cosh \frac{t}{3}-1}, \\
\sum\limits_{j=1}^{7}\left[ K^{\mathbf{E}_{7}}(t)\right] _{7j}^{-1}
&=&\sum\limits_{j=1}^{7}\left[ K^{\mathbf{E}_{7}}(t)\right] _{2j}^{-1}+\frac{%
1+2\left[ 2\sum_{k=1}^{3}\cosh \frac{kt}{18}-\cosh \frac{t}{6}\right] }{%
1-2\cosh \frac{t}{3}}.
\end{eqnarray}
Therefore 
\begin{eqnarray*}
\sum_{j,k,p=1}^{7}\left[ K^{\mathbf{E}_{7}}(t)\right] _{1j}^{-1}\chi
_{j}^{kp}\left[ K^{\mathbf{E}_{7}}(\frac{t}{2})\right] _{kp}^{-1} &=&\frac{%
4\left( \sum\limits_{k=2}^{3}\cosh \frac{kt}{18}+\cosh \frac{5t}{18}+\cosh 
\frac{7t}{18}\right) \left[ K^{\mathbf{E}_{7}}(\frac{t}{2})\right] _{11}^{-1}%
}{\left( 1-2\cosh \frac{t}{18}+2\cosh \frac{t}{9}\right) \left( 1-2\cosh 
\frac{t}{3}\right) }, \\
\sum_{j,k,p=1}^{7}\left[ K^{\mathbf{E}_{7}}(t)\right] _{2j}^{-1}\chi
_{j}^{kp}\left[ K^{\mathbf{E}_{7}}(\frac{t}{2})\right] _{kp}^{-1} &=&\frac{%
2-4\sum\limits_{k=2}^{5}\cosh \frac{kt}{18}+\cosh ^{-1}\frac{t}{18}}{2\cosh 
\frac{t}{3}-1}\left[ K^{\mathbf{E}_{7}}(\frac{t}{2})\right] _{22}^{-1}, \\
\sum_{j,k,p=1}^{7}\left[ K^{\mathbf{E}_{7}}(t)\right] _{3j}^{-1}\chi
_{j}^{kp}\left[ K^{\mathbf{E}_{7}}(\frac{t}{2})\right] _{kp}^{-1} &=&\frac{%
2\left( 1+2\cosh \frac{2t}{9}+2\cosh \frac{5t}{18}\right) }{2\cosh \frac{t}{3%
}-1}\left[ K^{\mathbf{E}_{7}}(\frac{t}{2})\right] _{33}^{-1},
\end{eqnarray*}

\begin{eqnarray*}
\sum_{j,k,p=1}^{7}\left[ K^{\mathbf{E}_{7}}(t)\right] _{4j}^{-1}\chi
_{j}^{kp}\left[ K^{\mathbf{E}_{7}}(\frac{t}{2})\right] _{kp}^{-1} &=&\frac{%
4\left( \cosh \frac{t}{18}+\cosh \frac{5t}{18}\right) -\cosh ^{-1}\frac{t}{18%
}}{2\cosh \frac{t}{3}-1}\left[ K^{\mathbf{E}_{7}}(\frac{t}{2})\right]
_{44}^{-1}, \\
\sum_{j,k,p=1}^{7}\left[ K^{\mathbf{E}_{7}}(t)\right] _{5j}^{-1}\chi
_{j}^{kp}\left[ K^{\mathbf{E}_{7}}(\frac{t}{2})\right] _{kp}^{-1} &=&\frac{%
\left[ 4\left( 1-\cosh \frac{t}{9}+\sum\limits_{k=4}^{5}\cosh \frac{kt}{18}%
\right) +\frac{1}{\cosh \frac{t}{18}}\right] \left[ K^{\mathbf{E}_{7}}(\frac{%
t}{2})\right] _{55}^{-1}}{1-2\cosh \frac{t}{3}},~~~~\ ~~ \\
\sum_{j,k,p=1}^{7}\left[ K^{\mathbf{E}_{7}}(t)\right] _{6j}^{-1}\chi
_{j}^{kp}\left[ K^{\mathbf{E}_{7}}(\frac{t}{2})\right] _{kp}^{-1} &=&4\frac{%
\sum\limits_{k=3}^{5}\cosh \frac{kt}{18}}{1-2\cosh \frac{t}{3}}\left[ K^{%
\mathbf{E}_{7}}(\frac{t}{2})\right] _{66}^{-1}, \\
\sum_{j,k,p=1}^{7}\left[ K^{\mathbf{E}_{7}}(t)\right] _{7j}^{-1}\chi
_{j}^{kp}\left[ K^{\mathbf{E}_{7}}(\frac{t}{2})\right] _{kp}^{-1} &=&\frac{%
\left[ 4\left( \sum\limits_{k=2}^{3}\cosh \frac{kt}{18}+\sum%
\limits_{k=6}^{7}\cosh \frac{kt}{18}\right) +\frac{1}{\cosh \frac{t}{18}}%
\right] \left[ K^{\mathbf{E}_{7}}(\frac{t}{2})\right] _{77}^{-1}}{\left(
1-2\cosh \frac{t}{9}\right) \left( 2\cosh \frac{t}{3}-1\right) }
\end{eqnarray*}
which suffices to compute $R.$

\subsection{$\mathbf{E}_{8}$}

Developing the determinant gives now the recursive equations 
\begin{eqnarray}
\det K^{\mathbf{E}_{8}} &=&\left[ K^{\mathbf{E}_{8}/\mathbf{A}_{1}}\right]
^{2}\left[ \det K^{\mathbf{E}_{8}/\mathbf{A}_{6}}-\det K^{\mathbf{E}_{8}/%
\mathbf{A}_{4}}\right] -\det K^{\mathbf{E}_{8}/\mathbf{A}_{6}}  \notag \\
&=&K^{\mathbf{E}_{8}/\mathbf{A}_{1}}\det K^{\mathbf{E}_{8}/\mathbf{D}%
_{7}}-\det K^{\mathbf{E}_{8}/\mathbf{A}_{6}}.
\end{eqnarray}
Again we can use the quantities already determined in the previous sections.
Specializing $K^{\mathbf{E}_{8}/\mathbf{A}_{1}}=2\cosh t/30$, we compute 
\begin{equation}
\left[ K^{\mathbf{E}_{8}}(t)\right] ^{-1}=\frac{1}{\mathbf{E}_{8}}\left( 
\begin{array}{cccccccc}
\mathbf{D}_{7} & \mathbf{A}_{4} & \mathbf{A}_{6} & \mathbf{A}_{1}\mathbf{A}%
_{4} & \mathbf{A}_{1}\mathbf{A}_{3} & \mathbf{A}_{1}\mathbf{A}_{2} & \mathbf{%
A}_{1}^{2} & \mathbf{A}_{1} \\ 
\mathbf{A}_{4} & \mathbf{A}_{7} & \mathbf{A}_{1}\mathbf{A}_{4} & \mathbf{A}%
_{2}\mathbf{A}_{4} & \mathbf{A}_{2}\mathbf{A}_{3} & \mathbf{A}_{2}^{2} & 
\mathbf{A}_{1}\mathbf{A}_{2} & \mathbf{A}_{2} \\ 
\mathbf{A}_{6} & \mathbf{A}_{1}\mathbf{A}_{4} & \mathbf{A}_{1}\mathbf{A}_{6}
& \mathbf{A}_{1}^{2}\mathbf{A}_{4} & \mathbf{A}_{1}^{2}\mathbf{A}_{3} & 
\mathbf{A}_{1}^{2}\mathbf{A}_{2} & \mathbf{A}_{1}^{3} & \mathbf{A}_{1}^{2}
\\ 
\mathbf{A}_{1}\mathbf{A}_{4} & \mathbf{A}_{2}\mathbf{A}_{4} & \mathbf{A}%
_{1}^{2}\mathbf{A}_{4} & \mathbf{A}_{1}\mathbf{A}_{2}\mathbf{A}_{4} & 
\mathbf{A}_{1}\mathbf{A}_{2}\mathbf{A}_{3} & \mathbf{A}_{1}\mathbf{A}_{2}^{2}
& \mathbf{A}_{1}^{2}\mathbf{A}_{2} & \mathbf{A}_{1}\mathbf{A}_{2} \\ 
\mathbf{A}_{1}\mathbf{A}_{3} & \mathbf{A}_{2}\mathbf{A}_{3} & \mathbf{A}%
_{1}^{2}\mathbf{A}_{3} & \mathbf{A}_{1}\mathbf{A}_{2}\mathbf{A}_{3} & 
\mathbf{A}_{3}\mathbf{A}_{4} & \mathbf{A}_{2}\mathbf{A}_{4} & \mathbf{A}_{1}%
\mathbf{A}_{4} & \mathbf{A}_{4} \\ 
\mathbf{A}_{1}\mathbf{A}_{3} & \mathbf{A}_{2}^{2} & \mathbf{A}_{1}^{2}%
\mathbf{A}_{2} & \mathbf{A}_{1}\mathbf{A}_{2}^{2} & \mathbf{A}_{2}\mathbf{A}%
_{4} & \mathbf{D}_{5}\mathbf{A}_{2} & \mathbf{D}_{5}\mathbf{A}_{1} & \mathbf{%
D}_{5} \\ 
\mathbf{A}_{1}^{2} & \mathbf{A}_{1}\mathbf{A}_{2} & \mathbf{A}_{1}^{3} & 
\mathbf{A}_{1}^{2}\mathbf{A}_{2} & \mathbf{A}_{1}\mathbf{A}_{4} & \mathbf{D}%
_{5}\mathbf{A}_{1} & \mathbf{A}_{1}\mathbf{E}_{6} & \mathbf{E}_{6} \\ 
\mathbf{A}_{1} & \mathbf{A}_{2} & \mathbf{A}_{1}^{2} & \mathbf{A}_{1}\mathbf{%
A}_{2} & \mathbf{A}_{4} & \mathbf{D}_{5} & \mathbf{E}_{6} & \mathbf{E}_{7}%
\end{array}
\right) .~~~~~
\end{equation}
We abbreviated here $g\equiv \det K^{\mathbf{E}_{8}/\mathbf{g}}$. The sum of
particular rows and column gives

\begin{eqnarray*}
\sum\limits_{j=1}^{8}\left[ K^{\mathbf{E}_{8}}(t)\right] _{1j}^{-1} &=&\frac{%
2\left( 3\left[ 1+\sum\limits_{k=3}^{4}\cosh \frac{kt}{30}\right]
+5\sum\limits_{k=1}^{2}\cosh \frac{kt}{30}+\sum\limits_{k=5}^{7}\cosh \frac{%
kt}{30}+\cosh \frac{t}{6}\right) }{2\cosh \frac{t}{5}+2\cosh \frac{4t}{15}%
-2\cosh \frac{t}{15}-1}, \\
\sum\limits_{j=1}^{8}\left[ K^{\mathbf{E}_{8}}(t)\right] _{2j}^{-1} &=&\frac{%
2\left( 4+\sum\limits_{k=1}^{3}(9-k)\cosh \frac{kt}{30}+\sum%
\limits_{k=4}^{5}(8-k)\cosh \frac{kt}{30}+\sum\limits_{k=6}^{7}\cosh \frac{kt%
}{30}\right) }{2\cosh \frac{t}{5}+2\cosh \frac{4t}{15}-2\cosh \frac{t}{15}-1}%
, \\
\sum\limits_{j=1}^{8}\left[ K^{\mathbf{E}_{8}}(t)\right] _{3j}^{-1}
&=&\sum\limits_{j=1}^{8}\left[ K^{\mathbf{E}_{8}}(t)\right] _{2j}^{-1}+\frac{%
3\left[ 1+2\cosh \frac{t}{30}\right] +4\sum\limits_{k=2}^{3}\cosh \frac{kt}{%
30}+2\sum\limits_{k=4}^{6}\cosh \frac{kt}{30}}{2\cosh \frac{t}{5}+2\cosh 
\frac{4t}{15}-2\cosh \frac{t}{15}-1}, \\
\sum\limits_{j=1}^{8}\left[ K^{\mathbf{E}_{8}}(t)\right] _{3j}^{-1}
&=&\sum\limits_{j=1}^{8}\left[ K^{\mathbf{E}_{8}}(t)\right] _{3j}^{-1}+\frac{%
4\cosh ^{2}\frac{t}{30}\left( 1+2\sum\limits_{k=1}^{4}\cosh \frac{kt}{30}%
+2\cosh \frac{t}{15}\right) }{2\cosh \frac{t}{5}+2\cosh \frac{4t}{15}-2\cosh 
\frac{t}{15}-1},
\end{eqnarray*}
\begin{eqnarray*}
\sum\limits_{j=1}^{8}\left[ K^{\mathbf{E}_{8}}(t)\right] _{5j}^{-1}
&=&\sum\limits_{j=1}^{8}\left[ K^{\mathbf{E}_{8}}(t)\right] _{3j}^{-1}+\frac{%
1+4\sum_{k=1}^{5}\cosh \frac{kt}{30}-2\cosh \frac{t}{6}}{2\cosh \frac{t}{5}%
+2\cosh \frac{4t}{15}-2\cosh \frac{t}{15}-1}, \\
\sum\limits_{j=1}^{8}\left[ K^{\mathbf{E}_{8}}(t)\right] _{6j}^{-1}
&=&\sum\limits_{j=1}^{8}\left[ K^{\mathbf{E}_{8}}(t)\right] _{3j}^{-1}+\frac{%
1+6\cosh \frac{t}{30}-2\left( \cosh \frac{t}{10}+\cosh \frac{2t}{15}\right) 
}{2\cosh \frac{t}{5}+2\cosh \frac{4t}{15}-2\cosh \frac{t}{15}-1}, \\
\sum\limits_{j=1}^{8}\left[ K^{\mathbf{E}_{8}}(t)\right] _{7j}^{-1}
&=&\sum\limits_{j=1}^{8}\left[ K^{\mathbf{E}_{8}}(t)\right] _{3j}^{-1}-\frac{%
2\left( 3+\sum_{k=1}^{5}(6-k)\cosh \frac{kt}{30}+\cosh \frac{2t}{15}\right) 
}{2\cosh \frac{t}{5}+2\cosh \frac{4t}{15}-2\cosh \frac{t}{15}-1}, \\
\sum\limits_{j=1}^{8}\left[ K^{\mathbf{E}_{8}}(t)\right] _{8j}^{-1}
&=&\sum\limits_{j=1}^{8}\left[ K^{\mathbf{E}_{8}}(t)\right] _{7j}^{-1}-\frac{%
4\cosh ^{2}\frac{t}{30}\left( 1+2\left[ \cosh \frac{t}{30}%
+\sum_{k=3}^{4}\cosh \frac{kt}{30}\right] \right) }{2\cosh \frac{t}{5}%
+2\cosh \frac{4t}{15}-2\cosh \frac{t}{15}-1}.
\end{eqnarray*}
Then we obtain 
\begin{eqnarray*}
\sum_{j,k,p=1}^{8}\left[ K^{\mathbf{E}_{8}}(t)\right] _{1j}^{-1}\chi
_{j}^{kp}\left[ K^{\mathbf{E}_{8}}(\frac{t}{2})\right] _{kp}^{-1} &=&\frac{%
2\left( 1+2\cosh \frac{t}{30}-2\sum\limits_{k=3}^{9}\cosh \frac{kt}{30}%
\right) }{-1+2\cosh \frac{t}{3}}\left[ K^{\mathbf{E}_{8}}(\frac{t}{2})\right]
_{11}^{-1}, \\
\sum_{j,k,p=1}^{8}\left[ K^{\mathbf{E}_{8}}(t)\right] _{2j}^{-1}\chi
_{j}^{kp}\left[ K^{\mathbf{E}_{8}}(\frac{t}{2})\right] _{kp}^{-1} &=&\frac{%
8\cosh \frac{t}{30}\left( 1+\cosh \frac{t}{30}\left[ 1-2\sum%
\limits_{k=1}^{2}\cosh \frac{kt}{15}\right] \right) \left[ K^{\mathbf{E}%
_{8}}(\frac{t}{2})\right] _{22}^{-1}}{2\cosh \frac{t}{5}+2\cosh \frac{4t}{15}%
-2\cosh \frac{t}{15}-1}, \\
\sum_{j,k,p=1}^{8}\left[ K^{\mathbf{E}_{8}}(t)\right] _{3j}^{-1}\chi
_{j}^{kp}\left[ K^{\mathbf{E}_{8}}(\frac{t}{2})\right] _{kp}^{-1} &=&\frac{%
4\left( \sum_{k=4}^{7}\cosh \frac{t}{6}-\cosh \frac{t}{30}\right) }{1+2\cosh 
\frac{t}{15}-2\cosh \frac{t}{5}-2\cosh \frac{4t}{15}}\left[ K^{\mathbf{E}%
_{8}}(\frac{t}{2})\right] _{33}^{-1}, \\
\sum_{j,k,p=1}^{8}\left[ K^{\mathbf{E}_{8}}(t)\right] _{4j}^{-1}\chi
_{j}^{kp}\left[ K^{\mathbf{E}_{8}}(\frac{t}{2})\right] _{kp}^{-1} &=&\frac{%
4\left( \cosh \frac{t}{6}+\cosh \frac{7t}{30}\right) }{1+2\cosh \frac{t}{15}%
-2\cosh \frac{t}{5}-2\cosh \frac{4t}{15}}\left[ K^{\mathbf{E}_{8}}(\frac{t}{2%
})\right] _{44}^{-1},
\end{eqnarray*}
\begin{eqnarray*}
\sum_{j,k,p=1}^{8}\left[ K^{\mathbf{E}_{8}}(t)\right] _{5j}^{-1}\chi
_{j}^{kp}\left[ K^{\mathbf{E}_{8}}(\frac{t}{2})\right] _{kp}^{-1} &=&\frac{%
2+4\left( \sum\limits_{k=5}^{7}\cosh \frac{kt}{30}-\cosh \frac{t}{10}\right) 
}{1+2\cosh \frac{t}{15}-2\cosh \frac{t}{5}-2\cosh \frac{4t}{15}}\left[ K^{%
\mathbf{E}_{8}}(\frac{t}{2})\right] _{55}^{-1}, \\
\sum_{j,k,p=1}^{8}\left[ K^{\mathbf{E}_{8}}(t)\right] _{6j}^{-1}\chi
_{j}^{kp}\left[ K^{\mathbf{E}_{8}}(\frac{t}{2})\right] _{kp}^{-1} &=&\frac{%
1+2\left( \sum\limits_{k=5}^{9}\cosh \frac{kt}{30}+2\cosh \frac{7t}{30}%
\right) }{\left( 1+2\cosh \frac{t}{15}-2\sum\limits_{k=3}^{4}\cosh \frac{kt}{%
15}\right) \cosh \frac{t}{15}}\left[ K^{\mathbf{E}_{8}}(\frac{t}{2})\right]
_{66}^{-1}, \\
\sum_{j,k,p=1}^{8}\left[ K^{\mathbf{E}_{8}}(t)\right] _{7j}^{-1}\chi
_{j}^{kp}\left[ K^{\mathbf{E}_{8}}(\frac{t}{2})\right] _{kp}^{-1} &=&\frac{%
2+4\sum\limits_{k=6}^{9}\cosh \frac{kt}{30}}{1-2\cosh \frac{t}{3}}\left[ K^{%
\mathbf{E}_{8}}(\frac{t}{2})\right] _{77}^{-1}, \\
\sum_{j,k,p=1}^{8}\left[ K^{\mathbf{E}_{8}}(t)\right] _{8j}^{-1}\chi
_{j}^{kp}\left[ K^{\mathbf{E}_{8}}(\frac{t}{2})\right] _{kp}^{-1} &=&\frac{%
4\left( \cosh \frac{t}{30}+\sum\limits_{k=6}^{10}\cosh \frac{kt}{30}\right) %
\left[ K^{\mathbf{E}_{8}}(\frac{t}{2})\right] _{88}^{-1}}{\left( 2\cosh 
\frac{t}{10}-1\right) \left( 1+2\cosh \frac{t}{15}-2\cosh \frac{t}{5}-2\cosh 
\frac{4t}{15}\right) },
\end{eqnarray*}
from which we can deduce directly $R$.

\bibliographystyle{phreport}
\bibliography{MatrixRef}

\begin{thebibliography}{10}

\bibitem{Toda}
A.~V. Mikhailov, M.~A. Olshanetsky, and A.~M. Perelomov,
\newblock Two dimensional generalized Toda lattice,
\newblock Commun. Math. Phys. {\bf 79}, 473 (1981).

\bibitem{Toda2}
D.~I. Olive and N.~Turok,
\newblock Local conserved densities and zero curvature conditions for Toda
  lattice field theories,
\newblock Nucl. Phys. {\bf B257}, 277 (1985).

\bibitem{FKS}
A.~Fring, C.~Korff, and B.~J. Schulz,
\newblock On the universal representation of the scattering matrix of affine
  Toda field theory,
\newblock Nucl. Phys. {\bf B567}, 409--453 (2000).

\bibitem{Oota:1997un}
T.~Oota,
\newblock q-deformed Coxeter element in non-simply laced affine Toda field
  theories,
\newblock Nucl. Phys. {\bf B504}, 738--752 (1997).

\bibitem{Yang}
C.-N. Yang,
\newblock Some exact results for the many body problems in one dimension with
  repulsive delta function interaction,
\newblock Phys. Rev. Lett. {\bf 19}, 1312--1314 (1967).

\bibitem{Baxter}
R.~J. Baxter,
\newblock One-dimensional anisotropic Heisenberg chain,
\newblock Annals Phys. {\bf 70}, 323--327 (1972).

\bibitem{Cherednik:1984vs}
I.~V. Cherednik,
\newblock Factorizing Particles on a half line and root systems,
\newblock Theor. Math. Phys. {\bf 61}, 977--983 (1984).

\bibitem{Sklyanin:1988yz}
E.~K. Sklyanin,
\newblock Boundary conditions for integrable quantum systems,
\newblock J. Phys. {\bf A21}, 2375 (1988).

\bibitem{Kulish}
P.~Kulish,
\newblock On universal solution to the reflection equation,
\newblock Talk presented at the conference on Infinite Dimensional Algebras and
  Quantum Integrable Systems  (Faro, Portugal, July 2003).

\bibitem{FK1}
A.~Fring and R.~K{\"o}berle,
\newblock Factorized scattering in the presence of reflecting boundaries,
\newblock Nucl. Phys. {\bf B421}, 159--172 (1994).

\bibitem{FK2}
A.~Fring and R.~K{\"o}berle,
\newblock Affine Toda field theory in the presence of reflecting boundaries,
\newblock Nucl. Phys. {\bf B419}, 647--664 (1994).

\bibitem{Sasaki:1993xr}
R.~Sasaki,
\newblock Reflection Bootstrap equations for Toda field theory,
\newblock Hangzhou Proceedings, Interface between Mathematics and Physics, eds.
  W. Nahm and J. Shen (World Scientific ) , hep--th/9311027 (1994).

\bibitem{FK3}
A.~Fring and R.~K{\"o}berle,
\newblock Boundary bound states in affine Toda field theory,
\newblock Int. J. Mod. Phys. {\bf A10}, 739--752 (1995).

\bibitem{Corrigan:1994ft}
E.~Corrigan, P.~E. Dorey, R.~H. Rietdijk, and R.~Sasaki,
\newblock Affine Toda field theory on a half line,
\newblock Phys. Lett. {\bf B333}, 83--91 (1994).

\bibitem{Kim:1995aq}
J.~D. Kim,
\newblock Boundary reflection matrix in perturbative quantum field theory,
\newblock Phys. Lett. {\bf B353}, 213--221 (1995).

\bibitem{Bowcock:1995vp}
P.~Bowcock, E.~Corrigan, P.~E. Dorey, and R.~H. Rietdijk,
\newblock Classically integrable boundary conditions for affine Toda field
  theories,
\newblock Nucl. Phys. {\bf B445}, 469--500 (1995).

\bibitem{Gandenberger:1995gg}
G.~M. Gandenberger and N.~J. MacKay,
\newblock Exact S matrices for $d_{(N+1)}^{(2)}$ affine Toda solitons and their
  bound states,
\newblock Nucl. Phys. {\bf B457}, 240--272 (1995).

\bibitem{Fujii:1995vc}
A.~Fujii and R.~Sasaki,
\newblock Boundary effects in integrable field theory on a half line,
\newblock Prog. Theor. Phys. {\bf 93}, 1123--1134 (1995).

\bibitem{Bowcock:1996gw}
P.~Bowcock, E.~Corrigan, and R.~H. Rietdijk,
\newblock Background field boundary conditions for affine Toda field theories,
\newblock Nucl. Phys. {\bf B465}, 350--364 (1996).

\bibitem{Penati:1996xp}
S.~Penati, A.~Refolli, and D.~Zanon,
\newblock Classical Versus Quantum Symmetries for Toda Theories with a
  Nontrivial Boundary Perturbation,
\newblock Nucl. Phys. {\bf B470}, 396--418 (1996).

\bibitem{Penati:1996js}
S.~Penati, A.~Refolli, and D.~Zanon,
\newblock Quantum boundary currents for nonsimply-laced Toda theories,
\newblock Phys. Lett. {\bf B369}, 16--22 (1996).

\bibitem{Kim:1996jq}
J.~D. Kim and I.~G. Koh,
\newblock Square Root Singularity in Boundary Reflection Matrix,
\newblock Phys. Lett. {\bf B388}, 550--556 (1996).

\bibitem{Delius:1998jw}
G.~W. Delius,
\newblock Restricting affine Toda theory to the half-line,
\newblock JHEP {\bf 09}, 016 (1998).

\bibitem{Delius:1998rf}
G.~W. Delius,
\newblock Soliton-preserving boundary condition in affine Toda field theories,
\newblock Phys. Lett. {\bf B444}, 217--223 (1998).

\bibitem{Dorey:1998kt}
P.~Dorey, R.~Tateo, and G.~Watts,
\newblock Generalisations of the Coleman-Thun mechanism and boundary reflection
  factors,
\newblock Phys. Lett. {\bf B448}, 249--256 (1999).

\bibitem{Gand1}
G.~M. Gandenberger,
\newblock On $a^{(2)}_{(1)}$ reflection matrices and affine Toda theories,
\newblock Nucl. Phys. {\bf B542}, 659--693 (1999).

\bibitem{Gand2}
G.~W. Delius and G.~M. Gandenberger,
\newblock Particle reflection amplitudes in $a^{(n)}_{(1)}$ Toda field
  theories,
\newblock Nucl. Phys. {\bf B554}, 325--364 (1999).

\bibitem{Bowcock1}
P.~Bowcock,
\newblock Classical backgrounds and scattering for affine Toda theory on a
  half-line,
\newblock JHEP {\bf 05}, 008 (1998).

\bibitem{Perkins}
M.~Perkins and P.~Bowcock,
\newblock Quantum corrections to the classical reflection factor in
  $a^{(2)}_{(1)}$ Toda field theory,
\newblock Nucl. Phys. {\bf B538}, 612--630 (1999).

\bibitem{Riva}
V.~Riva,
\newblock Boundary bootstrap principle in two-dimensional integrable quantum
  field theories,
\newblock Nucl. Phys. {\bf B604}, 511--536 (2001).

\bibitem{Ahn:2001fr}
C.~Ahn, C.~Kim, and C.~Rim,
\newblock Reflection amplitudes of boundary Toda theories and thermodynamic
  Bethe ansatz,
\newblock Nucl. Phys. {\bf B628}, 486--504 (2002).

\bibitem{Fateev:2001rk}
V.~A. Fateev,
\newblock Expectation values of boundary fields in integrable boundary Toda
  theories,
\newblock Mod. Phys. Lett. {\bf A16}, 1201--1212 (2001).

\bibitem{Delius:2001qh}
G.~W. Delius and N.~J. MacKay,
\newblock Quantum group symmetry in sine-Gordon and affine Toda field theories
  on the half-time,
\newblock Commun. Math. Phys. {\bf 233}, 173--190 (2003).

\bibitem{Fateev1}
V.~A. Fateev,
\newblock Normalization factors, reflection amplitudes and integrable systems,
\newblock hep-th/0103014  (2001).

\bibitem{Fateev2}
V.~A. Fateev and E.~Onofri,
\newblock Boundary one-point functions, scattering theory and vacuum solutions
  in integrable systems,
\newblock Nucl. Phys. {\bf B634}, 546--570 (2002).

\bibitem{Fateev3}
V.~A. Fateev and E.~Onofri,
\newblock Boundary one-point functions, scattering, and background vacuum
  solutions in Toda theories,
\newblock Int. J. Mod. Phys. {\bf A18}, 879--900 (2003).

\bibitem{Kojima:2002tc}
T.~Kojima,
\newblock The affine $A^{(1)}_{(n-1)}$ Toda fields with boundary reflection,
\newblock Int. J. Mod. Phys. {\bf A17}, 487--513 (2002).

\bibitem{Bowcock2}
P.~Bowcock and M.~Perkins,
\newblock Aspects of classical backgrounds and scattering for affine Toda
  theory on a half-line,
\newblock JHEP {\bf 02}, 016 (2003).

\bibitem{sinh}
S.~Ghoshal,
\newblock Bound state boundary S matrix of the Sine-Gordon model,
\newblock Int. J. Mod. Phys. {\bf A9}, 4801--4810 (1994).

\bibitem{sinh5}
E.~Corrigan,
\newblock On duality and reflection factors for the sinh-Gordon model with a
  boundary,
\newblock Int. J. Mod. Phys. {\bf A13}, 2709--2722 (1998).

\bibitem{sinh6}
H.~S. Cho, K.~S. Soh, and J.~D. Kim,
\newblock One-loop boundary reflection for the integrable boundary sinh-Gordon
  model,
\newblock J. Korean Phys. Soc. {\bf 32}, 661--665 (1998).

\bibitem{sinh4}
E.~Corrigan and G.~W. Delius,
\newblock Boundary breathers in the sinh-Gordon model,
\newblock J. Phys. {\bf A32}, 8601--8614 (1999).

\bibitem{sinh3}
A.~Chenaghlou and E.~Corrigan,
\newblock First order quantum corrections to the classical reflection factor of
  the sinh-Gordon model,
\newblock Int. J. Mod. Phys. {\bf A15}, 4417--4432 (2000).

\bibitem{sinh1}
E.~Corrigan and A.~Taormina,
\newblock Reflection factors and a two-parameter family of boundary bound
  states in the sinh-Gordon model,
\newblock J. Phys. {\bf A33}, 8739--8754 (2000).

\bibitem{sinh2}
M.~Ablikim and E.~Corrigan,
\newblock On the perturbative expansion of boundary reflection factors of the
  supersymmetric sinh-Gordon model,
\newblock Int. J. Mod. Phys. {\bf A16}, 625 (2001).

\bibitem{S13}
M.~Karowski and H.~J. Thun,
\newblock Complete S matrix of the massive Thirring model,
\newblock Nucl. Phys. {\bf B130}, 295 (1977).

\bibitem{ZZ}
A.~B. Zamolodchikov and A.~B. Zamolodchikov,
\newblock Factorized S-matrices in two dimensions as the exact solutions of
  certain relativistic quantum field models,
\newblock Annals Phys. {\bf 120}, 253--291 (1979).

\bibitem{S28}
A.~B. Zamolodchikov,
\newblock Exact two particel S matrix of quantum sine-Gordon solitons,
\newblock Pisma Zh. Eksp. Teor. Fiz. {\bf 25}, 499--502 (1977).

\bibitem{GhoshZ}
S.~Ghoshal and A.~B. Zamolodchikov,
\newblock Boundary S matrix and boundary state in two-dimensional integrable
  quantum field theory,
\newblock Int. J. Mod. Phys. {\bf A9}, 3841--3886 (1994).

\bibitem{Schroer:1976if}
B.~Schroer, T.~T. Truong, and P.~Weisz,
\newblock Towards an explicit construction of the sine-Gordon theory,
\newblock Phys. Lett. {\bf B63}, 422 (1976).

\bibitem{Karowski:1977th}
M.~Karowski, H.~J. Thun, T.~T. Truong, and P.~H. Weisz,
\newblock On the uniqueness of a purely elastic S matrix in (1+1) dimensions,
\newblock Phys. Lett. {\bf B67}, 321 (1977).

\bibitem{Zamolodchikov:1977uc}
A.~B. Zamolodchikov,
\newblock Exact S matrix of quantum sine-Gordon solitons,
\newblock JETP Lett. {\bf 25}, 468 (1977).

\bibitem{Dorey:1991xa}
P.~Dorey,
\newblock Root systems and purely elastic S matrices,
\newblock Nucl. Phys. {\bf B358}, 654--676 (1991).

\bibitem{Fring:1992gh}
A.~Fring and D.~I. Olive,
\newblock The Fusing rule and the scattering matrix of affine Toda theory,
\newblock Nucl. Phys. {\bf B379}, 429--447 (1992).

\bibitem{Kim:1995cf}
J.~D. Kim,
\newblock Boundary reflection matrix for A-D-E affine Toda field theory,
\newblock J. Phys. {\bf A29}, 2163--2174 (1996).

\bibitem{Kim}
J.~D. Kim,
\newblock Boundary reflection matrices for nonsimply laced affine Toda field
  theories,
\newblock Phys. Rev. {\bf D53}, 4441--4447 (1996).

\bibitem{Arinshtein:1979pb}
A.~E. Arinshtein, V.~A. Fateev, and A.~B. Zamolodchikov,
\newblock Quantum S matrix of the (1+1)-dimensional Todd chain,
\newblock Phys. Lett. {\bf B87}, 389--392 (1979).

\bibitem{Kim:1995qe}
J.~D. Kim and H.~S. Cho,
\newblock Boundary reflection matrix for $D_{4}^{(1)}$ affine Toda field
  theory,
\newblock hep-th/9505138  (1995).

\bibitem{Corrigan:1995np}
E.~Corrigan, P.~E. Dorey, and R.~H. Rietdijk,
\newblock Aspects of affine Toda field theory on a half line,
\newblock Prog. Theor. Phys. Suppl. {\bf 118}, 143--164 (1995).

\bibitem{sinhG1}
I.~Arefeva and V.~Korepin,
\newblock Scattering in two-dimensional model with Lagrangian
  $L=1/\gamma(1/2(\partial _mu u)^2+m^2(cos u-1))$,
\newblock Pisma Zh. Eksp. Teor. Fiz. {\bf 20}, 680 (1974).

\bibitem{sinhGS2}
S.~N. Vergeles and V.~M. Gryanik,
\newblock Two dimensional quantum field theories having exact solutions,
\newblock Yad. Fiz. {\bf 23}, 1324--1334 (1976).

\bibitem{Baseilhac:2002kf}
P.~Baseilhac and K.~Koizumi,
\newblock Sine-Gordon quantum field theory on the half-line with quantum
  boundary degrees of freedom,
\newblock Nucl. Phys. {\bf B649}, 491--510 (2003).

\bibitem{morecomingup}
O.~Castro-Alvaredo and A.~Fring,
\newblock Free parameters in boundary reflection matrices,
\newblock in preparation .

\bibitem{Fat2}
C.~Ahn, P.~Baseilhac, V.~A. Fateev, C.~Kim, and C.~Rim,
\newblock Reflection amplitudes in non-simply laced Toda theories and
  thermodynamic Bethe ansatz,
\newblock Phys. Lett. {\bf B481}, 114--124 (2000).

\bibitem{Fat3}
C.~Ahn, C.~Kim, and C.~Rim,
\newblock Reflection amplitudes of boundary Toda theories and thermodynamic
  Bethe ansatz,
\newblock Nucl. Phys. {\bf B628}, 486--504 (2002).

\bibitem{comingup}
O.~Castro-Alvaredo, A.~Fring, and M.~Stanishkov,
\newblock Boundary bootstrap equations versus classical boundary conditions,
\newblock in preparation .

\end{thebibliography}

\end{document}